\documentclass[prd,aps]{revtex4-2}

\usepackage{bm}
\usepackage[plainpages=false]{hyperref}
\usepackage{graphicx}
\usepackage{multirow}

\usepackage{color}
\usepackage[normalem]{ulem}

\begin{document}

\title{Simulations of Fast Neutrino Flavor Conversions with Interactions in Inhomogeneous Media}

\author{G\"unter Sigl$^{\it \bf 1}$\footnote[1]{E-mail: guenter.sigl@desy.de}}

\affiliation{
$^ {\bf \it 1}$Universit\"at Hamburg, {II}. Institute for Theoretical Physics, Luruper Chaussee 149, 22761 Hamburg, Germany
}

\begin{abstract}
We investigate toy models for spatial and temporal instabilities in collective neutrino oscillations induced by
neutrino self-interactions, with special emphasis on inhomogeneous systems with densities following a profile. 
Simulations are based on a mathematica program that solves the Liouville equation with or without vacuum terms, 
refractive terms from a background medium,
and neutrino-neutrino forward scattering, in one space dimension and in time. A discrete number of momentum 
modes are characterized by the neutrino velocity projection on the spatial direction. We also consider the effects of 
charged current interaction source terms and neutral current scattering contributions. We find that refractive effects 
from the medium, in particular for density distributions with a profile, and neutral current non-forward scattering
off the background medium can strongly influence fast collective flavor transformations. Specifically we find that
if both are present, fast flavor conversions can be strongly suppressed or at least delayed.
\end{abstract}

%\pacs{PACS numbers: }

\maketitle

\section{Introduction}
In environments with high neutrino densities the self-coupling of the known active electron-, muon- and tau-neutrinos leads
to interesting but at the same time complicated non-linear effects. For mass splittings $\Delta m^2$ of neutrinos of momentum
$p$ the interplay between vacuum oscillations with frequency $\omega=\Delta m^2/(2p)$
and self-coupling corresponding to a rate $\mu\sim \sqrt2 G_{\rm F}n_\nu$ can lead to so-called slow flavor
conversions, also known as bipolar pendulum-like oscillations~\cite{Hannestad:2006nj,Duan:2005cp} with rates $\sim(\omega\mu)^{1/2}$.
These kind of collective oscillations can also lead to a swapping of the energy spectra of electron-type neutrinos with
those of muon/tau-like neutrinos at a specific critical energy~\cite{Raffelt:2007xt,Raffelt:2007cb,Duan:2007bt,Fogli:2007bk,Martin:2019dof},
also known as spectral splits. Furthermore, it was pointed out by 
Sawyer~\cite{Sawyer:2005jk,Sawyer:2008zs} that small initial deviations from a pure flavor state can lead to so-called
fast flavor conversions proceeding with characteristic rates of order $\mu$. In the context of core collapse supernovae
around and inside the neutrinosphere where neutrino non-forward interactions decouple, one
typically has $\mu\gg\omega$ by several orders of magnitude such that fast flavor oscillations are in fact the most
efficient. Through a linear stability analysis it was later shown that this effect is driven by a flip in sign of the local
electron lepton minus muon/tau lepton number as function of the angle with respect to the radial direction (a so-called
flavor-lepton number crossing)~\cite{Izaguirre:2016gsx,Capozzi:2017gqd,Airen:2018nvp}. A particular difference between slow and
fast conversions is that
the basic version of the former can be described within ordinary differential equations in either time for a homogeneous system or in,
for example, radial direction, for a stationary system, whereas the latter depend on propagating flavor waves which thus
have to be described by partial differential equations. For some recent reviews see 
Ref.~\cite{Duan:2010bg,Chakraborty:2016yeg,Tamborra:2020cul}. The fast collective flavor oscillations in particular
are the subject of intense recent study, see e.g. Refs.~\cite{Martin:2019gxb,Bhattacharyya:2020jpj,Abbar:2019zoq,Abbar:2020ror}.

More recently, in attempts to make the analysis more and more realistic, more ingredients have been added and
some simplifying assumptions such as certain symmetries have been dropped. For example, bipolar oscillations can be
modified when the assumption of homogeneity is dropped and the system is described by partial differential 
equations~\cite{Mangano:2014zda}. While frequent non-forward scattering tends to damp neutrino oscillations~\cite{Stodolsky:1986dx}
neutrino scattering off the ambient matter can lead to a neutrino halo even outside the neutrino sphere where non-forward
scattering is rare. The influence of such a neutrino halo on the neutrino self-interactions
has been investigated in Refs.~\cite{Cherry:2012zw,Sarikas:2012vb}. The role of the convective terms in fast flavor
conversions with inhomogeneous initial conditions with otherwise spatially homogeneous couplings have been investigated in 
Ref.~\cite{Shalgar:2019qwg}. In addition, the role of charged current
source and sink terms for the neutrinos~\cite{Cirigliano:2017hmk,Capozzi:2018clo} as well as of non-forward neutral current collisions
in fast collective oscillations have been investigated~\cite{Shalgar:2020wcx,Martin:2021xyl}. Ref.~\cite{Johns:2021qby} found that
in the presence of a sufficiently large asymmetry between interaction rates for neutrinos ans anti-neutrinos a new kind
of collisional instability can occur. Three-flavor effects have also been considered~\cite{Shalgar:2021wlj}.
It turns out that most if not all of those ingredients can significantly modify the character of resulting flavor conversions.

In the current study we do not attempt to perform a complete treatment in any sense, but to develop a numerical toy
model setup that is at the same time simple enough to be run with reasonable resources and at the same time still
sufficiently complex to study some of the effects mentioned above at least in a qualitative way. To this end we will
solve a partial differential equation in one time and one (radial) dimension with a discrete number of momentum modes
that has a Liouville-type transport term on
the left hand side and a commutator describing vacuum, matter and collective oscillations as well as a collision term
involving charged and neutral current term on the right hand side.

To mimic the situation in a core collapse supernova we then numerically solve these equations in particular in the
context of radial profiles for the various rates. We investigate specifically the role of the matter oscillation term
and the collision terms in the presence of profiles.

In section~\ref{sec:sec1} we describe the relevant general partial differential equation in 3 space and one time dimension
with a continuum of momentum modes. In section~\ref{sec:sec2} we simplify this general equation to
one time and one (radial) dimension with discrete momentum modes. In section~\ref{sec:sec3} we then perform numerical
simulations with this equation and present results for some cases of interest. Section~\ref{sec:sec4} contains a discussion
of the results and we conclude in section~\ref{sec:sec5}.

\section{Kinetic Equations for Collective Oscillations} \label{sec:sec1}
We consider $N_f$ flavors of neutrinos and anti-neutrinos with annihilation operators $a_i$ and $b_i$
and creation operators $a^\dagger_i$ and $b^\dagger_i$, respectively, which act at a given location
${\bf r}$ or momentum ${\bf p}$. As dynamical variables we then use the corresponding Wigner distributions defined by
\begin{eqnarray}
  \rho_{ij}({\bf r},{\bf p})&\equiv&\int d^3{\bf r}^\prime\,e^{-i{\bf p}
  \cdot{\bf r}^\prime}\left\langle a^\dagger_j({\bf r}-{\bf r}^\prime/2)
  a_i({\bf r}+{\bf r}^\prime/2)\right\rangle\nonumber\\
  &=&\int {d^3{\bf\Delta}\over (2\pi)^3}\,e^{i{\bf\Delta}\cdot{\bf r}}
  \left\langle a^\dagger_j({\bf p}-{\bf\Delta}/2)a_i({\bf p}+
  {\bf\Delta}/2)\right\rangle\,,\\
  \bar\rho_{ij}({\bf r},{\bf p})&\equiv&\int d^3{\bf r}^\prime\,e^{-i{\bf p}
  \cdot{\bf r}^\prime}\left\langle b^\dagger_i({\bf r}-{\bf r}^\prime/2)
  b_j({\bf r}+{\bf r}^\prime/2)\right\rangle\nonumber\\
  &=&\int {d^3{\bf\Delta}\over (2\pi)^3}\,e^{i{\bf\Delta}\cdot{\bf r}}
  \left\langle b^\dagger_i({\bf p}-{\bf\Delta}/2)b_j({\bf p}+
  {\bf\Delta}/2)\right\rangle\,,\nonumber
\label{eq:Wigner}
\end{eqnarray}
see, e.g. Ref.~\cite{Stirner:2018ojk}. For a derivation and general discussion of the following kinetic equations see
e.g. Ref.~\cite{Sigl:1993ctk}. We consider the equations of motion for these variables,
\begin{eqnarray}\label{eq:eom1}
 \partial_t \rho({\bf r},{\bf p})+{\bf v}({\bf r},{\bf p})\cdot\nabla_{\bf r}\rho({\bf r},{\bf p})&=&
 -i\left[\Omega^0_{\bf p}+\Omega_m({\bf r})+\Omega^{\rm S}({\bf r},{\bf p})
 ,\rho_{\bf p}\right]+\partial_t \rho({\bf r},{\bf p})_{\rm coll}\,,\\
 \partial_t \bar\rho({\bf r},{\bf p})+{\bf v}({\bf r},{\bf p})\cdot\nabla_{\bf r}\bar\rho({\bf r},{\bf p})&=&
 +i\left[\Omega^0_{\bf p}-\Omega_m({\bf r})-\Omega^{\rm S}({\bf r},{\bf p})
 ,\bar\rho_{\bf p}\right]+\partial_t\bar\rho({\bf r},{\bf p})_{\rm coll}\,,\nonumber
\end{eqnarray}
where $[{\cdot},{\cdot}]$ is the commutator,
\begin{equation}\label{eq:Omega_m}
\Omega_m\equiv{\rm diag}[\lambda_1({\bf r}),\cdots,\lambda_n({\bf r})]\,,
\end{equation}
is the in general space-dependent background
matter contribution to the rotation frequency matrix which is diagonal in the flavor basis,
\begin{equation}\label{eq:Omega_p}
\Omega^0_{\bf p}\equiv\frac{1}{2p}\,{\rm diag}(m_1^2,\cdots,m_n^2)\,,
\end{equation}
is the matrix of vacuum oscillation frequencies, expressed in the mass basis, for
ultra-relativistic neutrinos, with $p=|{\bf p}|$, and the self-interactions are characterised by
\begin{equation}
\Omega^{\rm S}({\bf r},{\bf p})=\sqrt2 G_{\rm F}\int \frac{d^3{\bf q}}{(2\pi)^3}g_{{\bf p},{\bf q}}\left\{
    G_{\rm S}[\rho({\bf r},{\bf q})-\bar\rho({\bf r},{\bf q})]G_{\rm S}+G_{\rm S}
    {\rm Tr}\left[(\rho({\bf r},{\bf q})-\bar\rho({\bf r},{\bf q}))G_{\rm S}\right]\right\}\,,\label{eq:Onu}
\end{equation}
where $G_{\rm F}$ is Fermi's constant, $G_{\rm S}$ is a dimensionless hermitian matrix of coupling constants, which is
just the unit matrix for active Standard Model neutrinos, and $g_{{\bf p},{\bf q}}$
are dimensionless momentum mode dependent coupling constants. With ${\bf v}_{\bf p}$ the neutrino velocity in momentum
mode ${\bf p}$ it is generally given by $g_{{\bf p},{\bf q}}=1-{\bf v}_{\bf p}\cdot{\bf v}_{\bf q}$.

Furthermore, we schematically
added a collision term $\partial_t \rho({\bf r},{\bf p})_{\rm coll}$ on the r.h.s. of Eq.~(\ref{eq:eom1}).
It can have contributions from charged current source/sink terms of the form
\begin{eqnarray}
\partial_t \rho({\bf r},{\bf p})_{\rm coll,CC}&=&\left\{{\cal P}({\bf r},{\bf p}),\left(1-{\rho({\bf r},{\bf p})
  \over f_0({\bf r},{\bf p})}\right)\right\}\,,\nonumber\\
\partial_t{\bar\rho({\bf r},{\bf p})}_{\rm coll,CC}&=&\left\{{\cal A}({\bf r},{\bf p}),
  \left(1-{\bar\rho({\bf r},{\bf p})\over\bar
  f_0({\bf r},{\bf p})}\right)\right\}\,,\label{eq:kin_cc}
\end{eqnarray}
with $f_0({\bf r},{\bf p})$ and $\bar f_0({\bf r},{\bf p})$ equilibrium occupation numbers in mode ${\bf p}$
and ${\cal P}({\bf r},{\bf p})$ and ${\cal A}({\bf r},{\bf p})$ some matrix-valued rates that we will specify later.
Further contributions to the collision term can come from
neutral current interactions which can be written as
\begin{eqnarray}
\partial_t\rho({\bf r},{\bf p})_{\rm coll,NC}&=&
  {1\over 2}\int\frac{d^3{\bf q}}{(2\pi)^3}\biggl\{W({\bf r},q,p)
    (1-\rho_{\bf p})G\rho_{\bf q}G-W({\bf r},p,q)
    \rho_{\bf p}G(1-\rho_{\bf q})G\nonumber\\
  &&\hskip-5em+W({\bf r},-q,p)(1-\rho_{\bf p})
  G(1-\overline\rho_{\bf q})G
    -W({\bf r},p,-q)\rho_{\bf p}G\overline\rho_{\bf q}G+{\rm h.c.}
    \biggr\}\label{eq:kin_nc}\,,\\
\partial_t\overline\rho({\bf r},{\bf p})_{\rm coll,NC}&=&
  {1\over 2}\int\frac{d^3{\bf q}}{(2\pi)^3}\biggl\{W({\bf r},-p,-q)
    (1-\overline\rho_{\bf p})G\overline\rho_{\bf q}G-
    W({\bf r},-q,-p)\overline\rho_{\bf p}G(1-\overline\rho_{\bf q})G\nonumber\\
  &&\hskip-5em+W({\bf r},-p,q)(1-\overline\rho_{\bf p})G
    (1-\rho_{\bf q})G-
    W({\bf r},q,-p)\overline\rho_{\bf p}G\rho_{\bf q}G+{\rm h.c.}
    \biggr\}\,,\nonumber
\end{eqnarray}
where $q$ and $p$ are neutrino four momenta with positive energy $p_0=\vert{\bf p}\vert$ and $q_0=\vert{\bf q}\vert$,
$W({\bf r},q,p)$ are non-negative scattering rates per unit neutrino density and $G$ is another dimensionless hermitian matrix of coupling constants.

Generally the velocity of the neutrino ${\bf v}({\bf r},{\bf p})$ depends on ${\bf r}$ and ${\bf p}$. Strictly speaking,
the Liouville term on the left hand side of Eq.~(\ref{eq:eom1}) should read
\begin{equation}
  {1\over 2}\left\{\nabla_{\bf r}\rho({\bf r},{\bf p}),\nabla_{\bf p}
  \Omega({\bf r},{\bf p})\right\}-
  {1\over 2}\left\{\nabla_{\bf p}\rho({\bf r},{\bf p}),\nabla_{\bf r}
  \Omega({\bf r},{\bf p})\right\}\label{eq:drift_op}
\end{equation}
and similarly for $\dot{\bar\rho}({\bf r},{\bf p})$, where $\left\{.,.\right\}$ is the anticommutator.
Here, $\Omega({\bf r},{\bf p})$ is the matrix of total energies which
includes external potentials like gravity as well as the refractive\index{refraction}
energy shifts discussed above. The latter are represented by the
terms $\Omega^0_{\bf p}\pm\Omega_m({\bf r})\pm\Omega^{\rm S}({\bf r})$ in the commutators in Eq.~(\ref{eq:eom1}).
Therefore, the left hand side of Eq.~(\ref{eq:eom1}) would include a drift term proportional to the
neutrino velocity ${\bf v}=\nabla_{\bf p}\Omega({\bf r},{\bf p})$ and a term
proportional to the force ${\bf F}=-\nabla_{\bf r}\Omega({\bf r},{\bf p})$
acting onto the neutrino. However, we here neglect the force ${\bf F}$ and approximate
${\bf v}({\bf r},{\bf p})\simeq{\bf p}/E_{\bf p}$ and project it on a direction of interest.
Furthermore, the velocities ${\bf v}_{\bf p}$ and the couplings $g_{{\bf p},{\bf q}}=1-{\bf v}_{\bf p}\cdot{\bf v}_{\bf q}$
could be made radius or time dependent to mimic the geometry of a supernova, for example.

The initial conditions can be parametrized as
\begin{equation}\label{eq:initial2}
  \rho(t=0,{\bf r},{\bf p})=\frac{n({\bf r},{\bf p})}{2}\left(\matrix{1+\cos\theta({\bf r},{\bf p})&
  \exp[i\phi({\bf r},{\bf p})]\sin\theta({\bf r},{\bf p})\cr \exp[-i\phi({\bf r},{\bf p})]\sin\theta({\bf r},{\bf p})& 1-\cos\theta({\bf r},{\bf p})}\right)\,,
\end{equation}
where the total occupation numbers $n({\bf r},{\bf p})$ are fixed during evolution if non-forward scattering can be neglected
and the initial polarisations are characterised by the angles
$\theta({\bf r},{\bf p})$ and $\phi({\bf r},{\bf p})$. An analogous expression can be written for the initial anti-neutrino
matrices.

A unique solution of the partial differential equations also requires to specify boundary conditions. We will adapt those
to the specific problem we will consider in the next section.

\section{Numerical Implementation}\label{sec:sec2}

In general we use $N_f=2$ flavors, one spatial dimension $x$ and a fixed number $N_p$ of momentum modes
and thus characterise $\rho({\bf r},{\bf p})$
by $\rho(x,i_p)$ with a real spatial coordinate $x$ in the range $0\leq x\leq L$ and an integer $i_p$ in the range
$1\leq i_p\leq N_p$. Furthermore, we set $G_{\rm S}={\rm diag}(1,\ldots,1)$.

The $N_p$ momentum modes cover velocities in the unique direction $x$
within a range $v_{\rm min}\leq v_x(i_p)\leq v_{\rm max}$. All modes are assumed to have the same absolute value for the momentum, i.e.
we only consider angular modes at a given energy.

For the vacuum oscillation terms in Eq.~(\ref{eq:eom1}) we choose
\begin{equation}\label{eq:Omega_p2}
  \Omega^0_{p}=\frac{\Delta m^2}{4p}\left(\matrix{\cos2\theta_0& -\sin2\theta_0\cr -\sin2\theta_0& -\cos2\theta_0}\right)\,,
\end{equation}
in the flavor basis where $\theta_0$ is the vacuum mixing angle and $\Delta m^2=m_1^2-m_2^2$. Thus, for $\theta_0<\pi/4$
the inverted mass hierarchy corresponds to $\Delta m^2>0$.
In the cases we will consider there is no momentum dependence because constant absolute momentum is considered.
For the matter term we take
\begin{equation}\label{eq:Omega_l}
  \Omega_m(x)=\lambda(x)\sigma_3=\lambda(x)\left(\matrix{1& 0\cr 0& 0}\right)\,,
\end{equation}
with a scalar function $\lambda(x)$ that represents a rate that may depend on $x$.

To discretize the forward-scattering self-interaction term Eq.~(\ref{eq:Onu}) is a bit more tricky. In principle one would
substitute the momentum integral by a discrete sum $V^{-1}\sum_{\bf q}$ where $V$ is a quantization volume. Since
we here effectively reduce the problem to a one-dimensional one and only take into account a small number of radial
modes, we substitute $\sqrt2 G_{\rm F}V^{-1}\sum_{\bf q}g_{{\bf p},{\bf q}}$ by $\mu(x)\sum_{i_q}g_{i_p,i_q}$. Here,
$\mu(x)$ is an effective self-interaction rate of order $\sqrt2 G_{\rm F}n_\nu(x)$ with $n_\nu(x)$ the neutrino density.
The decrease of $\mu(x)$ with increasing $x$ mimics the fact that in three dimensions, due to dilution in the transverse
directions the neutrino density falls off faster than in the one-dimensional transport model used here.
The coupling constants $g_{i_p,i_q}$ are normalized such that the average of $\sum_{i_q}g_{i_p,i_q}$ over $i_p$ is unity.
For the toy scenarios below we use
\begin{equation}\label{eq:coupling}
  g_{i_p,j_q}=\frac{(1-\delta_{i_pj_q})[1-v_x(i_p)v_x(j_q)]}{\sum_{i_kj_l}(1-\delta_{i_kj_l})[1-v_x(i_k)v_x(j_l)]/N_p}\,,
\end{equation}
which assures that a given mode does not couple to itself and that the average coupling of one momentum mode
summed over all other modes in Eq.~(\ref{eq:Onu}) is unity and thus does not depend on $N_p$. With this we finally get
\begin{equation}
\Omega^{\rm S}(x,i_p)=\mu(x)\sum_{i_q\neq i_p}g_{i_p,i_q}\left\{
    G_{\rm S}[\rho(x,i_q)-\bar\rho(x,i_q)]G_{\rm S}+G_{\rm S}
    {\rm Tr}\left[(\rho(x,i_q)-\bar\rho(x,i_q))G_{\rm S}\right]\right\}\,.\label{eq:Onu2}
\end{equation}

For the charged current interaction term in one spatial dimension we use
\begin{eqnarray}
\partial_t \rho(x,i_p)_{\rm coll,CC}&=&f_{\rm CC}(x,i_p)\left\{\left(\matrix{1& 0\cr 0& 0}\right),\left(1-{\rho(x,i_p)
  \over f_0(x,i_p)}\right)\right\}\,,\nonumber\\
\partial_t{\bar\rho}(x,i_p)_{\rm coll,CC}&=&\bar f_{\rm CC}(x,i_p)\left\{\left(\matrix{1& 0\cr 0& 0}\right),
  \left(1-{\bar\rho(x,i_p)\over \bar f_0(x,i_p)}\right)\right\}\,,\label{eq:kin_cc2}
\end{eqnarray}
where $f_{\rm CC}(x,i_p)$, $\bar f_{\rm CC}(x,i_p)$, $f_0(x,i_p)$ and $\bar f_0(x,i_p)$ are suitably chosen functions. Eq.~(\ref{eq:kin_cc2}) describes the injection of a pure flavor with a rate characterised by $f_{\rm CC}(x,i_p)$ and $\bar f_{\rm CC}(x,i_p)$ and equilibrium occupation numbers $f_0(x,i_p)$ and
$\bar f_0(x,i_p)$.

For simplicity for the neutral current interactions
we here neglect pair creation and annihilation terms out of and into the medium and assume that the rates
$W({\bf r},q,p)$ do not depend on $q$ or $p$
(isotropic, energy-independent scattering). Then the terms quadratic in $\rho$ cancel and one can write
\begin{equation}\label{eq:kin_nc2}
\partial_t\rho(x,i_p)_{\rm coll,NC}=f_{\rm NC}(x)\left\{
  \frac{1}{N_p}\sum_{i_q}G\rho(x,i_q)G-\frac{1}{2}\left[\rho(x,i_p)G^2+G^2\rho(x,i_p)\right]\right\}\,,
\end{equation}
where $f_{\rm NC}(x)$ is a location dependent scattering rate and we use $G=G_{\rm S}$. An analogous equation
holds for anti-neutrinos. The scattering term is again normalized such that it does not depend on $N_p$.

Assuming isotropic energy-independent rates it is easy to see from Eq.~(\ref{eq:kin_nc}) that one can include neutrino
pair creation and pair annihilation out of and into the medium by generalizing Eq.~(\ref{eq:kin_nc2}) to
\begin{equation}\label{eq:kin_nc3}
\partial_t\rho(x,i_p)_{\rm coll,NC}=f_{\rm NC}(x)\left\{
   \frac{1}{N_p}\sum_{i_q}G\left[\rho(x,i_q)-\bar\rho(x,i_q)\right]G-\left[\rho(x,i_p)G^2+G^2\rho(x,i_p)\right]+G^2\right\}\,,
\end{equation}
with an analogous equation for anti-neutrinos. This would, however, be only realistic for neutrino energies much smaller
than the medium temperature because otherwise pair creation rates out of the medium would be thermally suppressed
with respect to pair annihilation into the medium, which calls for a more detailed implementation which we postpone to future work.
We will, therefore, use Eq.~(\ref{eq:kin_nc2}) throughout the present studies.

To model anisotropies and asymmetries between neutrinos and anti-neutrinos, let us now define
\begin{equation}\label{eq:aniso}
  h(x,i_p)=\left(2\frac{i_p-1}{N_p-1}-1\right)g(x)\,,
\end{equation}
where $g(x)$ vanishes at the boundaries, $g(0)=g(L)=0$. With this we also define the modulation factors
\begin{eqnarray}
f(x,i_p)&=&\frac{1}{2}[1-ah(x,i_p)][1-bh(x,i_p)]\,,\nonumber\\
\bar f(x,i_p)&=&\frac{1}{2}[1-ah(x,i_p)][1+bh(x,i_p)]\,,\label{eq:aniso2}
\end{eqnarray}
where $a$ and $b$ characterise the anisotropy in the neutrino plus anti-neutrino distribution $f(x,i_p)+\bar f(x,i_p)$ and the
asymmetry between neutrinos and anti-neutrinos, respectively. Thus, since $h(x,i_p)$ vanishes at the spatial boundaries by
construction, independent of the values for $a$ and $b$
we assume isotropy of the distribution at the boundaries within the chosen momentum range. Also note that
$a=b=0$ corresponds to an isotropic distribution everywhere in which case $h(x,i_p)$ is irrelevant.
Based on these definitions we now set equilibrium occupation numbers, injection rates and initial conditions proportional to
$f(x,i_p)$ and $\bar f(x,i_p)$ with $x-$dependent normalizations that can be suitably chosen. The equilibrium occupation numbers then become
\begin{equation}\label{eq:f_eq}
f_0(x,i_p)=f_0(x)f(x,i_p)\,,\quad\bar f_0(x,i_p)=f_0(x)\bar f(x,i_p)\,,
\end{equation}
where $f_0(x)$ represents an overall normalization, and the injection rates are
\begin{equation}\label{eq:f_s}
f_{\rm CC}(x,i_p)=f_s(x)f(x,i_p)\,,\quad\bar f_{\rm CC}(x,i_p)=f_s(x)\bar f(x,i_p)\,,
\end{equation}
where $f_s(x)$ again represents an overall normalization. Note that this ansatz ensures that any initial anisotroy and/or
neutrino/anti-neutrino asymmetry is supported by these source terms.

For the initial condition Eq.~(\ref{eq:initial2}) we use
\begin{equation}\label{eq:M_x3}
  \rho(t=0,x,i_p)=f_i(x)f(x,i_p)M(x)\,,\quad
  \bar\rho(t=0,x,i_p)=f_i(x)\bar f(x,i_p)M(x)\,,
\end{equation}
where the $x-$dependent $f_i(x)$ characterises the total neutrino plus anti-neutrino occupation number and the matrix $M$
is parametrized as
\begin{equation}\label{eq:M_x1}
  M(x)=\frac{1}{2}\left(\matrix{1+\cos\theta(x)& \exp[i\phi(x)]\sin\theta(x)\cr \exp[-i\phi(x)]\sin\theta(x)& 1-\cos\theta(x)}\right)\,.
\end{equation}
We will typically use random numbers for the angles $\theta(x)$ and $\phi(x)$, with a characteristic (small) amplitude $A$
for $\theta(x)$. This assures that initially the flavor state will be close to the dominant flavor 1, which in the supernova context
typically are electron neutrinos.

The ansatz Eq.~(\ref{eq:M_x3}) with Eq.~(\ref{eq:aniso2}) assures that for $b\neq0$ the lepton number of flavor 1
exhibits a flip of sign within the momentum range simulated
or, more generally, the lepton number asymmetry as function of the unit vector ${\bf n}$ characterising the direction
\begin{equation}\label{eq:ELN}
  G({\bf r},{\bf n})=\int_0^\infty\frac{dpp^2}{2\pi^2}\left[\rho_{11}({\bf r},p{\bf n})-\bar\rho_{11}({\bf r},p{\bf n})
  -\rho_{22}({\bf r},p{\bf n})+\bar\rho_{22}({\bf r},p{\bf n})\right]\,,
\end{equation}
changes sign as a function of ${\bf n}$.
This is known to be the criterion for fast flavor conversions to occur~\cite{Izaguirre:2016gsx,Dasgupta:2021gfs}.

Note that the ansätze above imply that there is no global asymmetry between neutrinos and anti-neutrinos. The parameter
$b$ in Eq.~(\ref{eq:aniso2}) only leads to a local asymmetry in momentum which averages out when summing over all momenta.
One could of course generalize this to a global asymmetry between neutrinos and anti-neutrinos by choosing some or
all of the normalizations $f_0(x)$, $f_s(x)$ and $f_i(x)$ different for neutrinos and anti-neutrinos.

\section{Results for Specific Toy Models} \label{sec:sec3}
We here have in mind the situation of a core collapse supernova in which $x$ represents the radial coordinate and
the inner boundary at $x=0$ lies inside the neutrino decoupling sphere at high densities, whereas the outer boundary at $x=L$
lies outside the neutrino sphere at low densities.
We apply angular modes covering $-1\leq v_x\leq1$, with central values
\begin{equation}\label{eq:v_x}
v_x(i_p)=-1+\frac{1}{N_p}+\frac{i_p-1}{N_p-1}\left(2-\frac{2}{N_p}\right)\,,\quad i_p=1,\cdots,N_p\,,
\end{equation}
with $N_p$ even. In the following we will sometimes label the momentum modes with $v_x$ instead of with $i_p$.
At the outer boundary at $x=L$ the boundary condition for the incoming modes, $v_x<0$, is then simply chosen as fixed by
the initial condition Eq.~(\ref{eq:M_x3}),
\begin{eqnarray}
  \rho(t,x=L,v_x<0)&=&\rho(t=0,x=L,v_x<0)\label{eq:M_init2a}\\
  \bar\rho(t,x=L,v_x<0)&=&\bar\rho(t=0,x=L,v_x<0)\,.\nonumber
\end{eqnarray}
The rationale is that there are very few incoming neutrinos at the outer boundary due to rare scattering
and they are of the dominant flavor 1. The boundary conditions at $x=0$ are instead given in terms of a reflective boundary,
\begin{eqnarray}\label{eq:M_init2b}
  \rho(t,x=0,v_x)=\rho(t,x=0,-v_x)\,,\quad\bar\rho(t,x=0,v_x)=\bar\rho(t,x=0,-v_x)\,.
\end{eqnarray}
Since at $x=0$ there is no momentum dependence for the initial conditions Eq.~(\ref{eq:M_x3}) because $f(x=0,i_p)=g(x=0)=0$,
this is consistent with the initial conditions for arbitrary values of $a$ and $b$. Here the rationale is that in the high density region
the distribution should be essentially isotropic due to frequent scattering.

In our simulations the total number of neutrinos at a given location and time is then given by
\begin{equation}\label{eq:N_tot}
  N(t,x)\equiv\sum_{v_x}{\rm Tr}\left[\rho(x,v_x)+\bar\rho(x,v_x)\right]\,.
\end{equation}
Note that Eqs.~(\ref{eq:aniso2}) and~(\ref{eq:M_x3}) imply that $N(t=0,x)=N_pf_i(x)$ initially because $\sum^{N_p}_{i_p=1}h(x,i_p)=0$
for $h(i_p)$ given by Eq.~(\ref{eq:aniso}).
With Eq.~(\ref{eq:N_tot}) we also define the normalized off-diagonal elements as
\begin{equation}\label{eq:F_off}
  F_{\rm off}(t,x)\equiv\frac{\sum_{v_x}\left|\rho_{12}(x,v_x)+\bar\rho_{12}(x,v_x)\right|}{N(t,x)}\,,
\end{equation}
and the normalized flavor asymmetry of the outgoing neutrino modes as
\begin{equation}\label{eq:F_asym}
  F_{\rm asym}(t,x)\equiv\frac{\sum_{v_x>0}\left[\rho_{11}(x,v_x)-\rho_{22}(x,v_x)+\bar\rho_{11}(x,v_x)-\bar\rho_{22}(x,v_x)\right]}
  {\sum_{v_x>0}{\rm Tr}\left[\rho(x,v_x)+\bar\rho(x,v_x)\right]}\,.
\end{equation}
We here restrict to the outgoing modes because those are the ones that in the end are observable and their occupation is
not fixed by the boundary condition at the outer boundary.

Apart from the length $L$ of the simulated range in $x$ we specify a length scale $x_0$ on which the $x-$dependent functions
in our problem vary. For the $x-$dependence of the modulation in
Eqs.~(\ref{eq:aniso}) and~(\ref{eq:aniso2}) we then generally use $g(x)=\sin(\pi x/L)\exp(-x/x_0)$, assuring $g(0)=g(L)=0$.
For the normalization of the equilibrium occupation numbers in Eq.~(\ref{eq:f_eq}) we choose $f_0(x)=\bar f_0(x)=0.8\exp(-x/x_0)$,
as well as for the normalization $f_i(x)=0.8\exp(-x/x_0)$ of the initial occupation numbers in Eq.~(\ref{eq:M_x3}).
For the normalization of the charged current rates in Eq.~(\ref{eq:f_s}) we choose $f_s(x)=0.1\exp(-x/x_0)$.
The same exponential profiles are used for the matter and self-interaction terms, $\lambda(x)=\lambda_0\exp(-x/x_0)$,
$\mu(x)=\mu_0\exp(-x/x_0)$.

Since we will here concentrate on fast flavor oscillations which do not depend on the vacuum term
and the vacuum term in a type II supernova is generally
much smaller than the matter and self-interaction terms, we here neglect it by putting $\Omega^0_{\bf p}=0$ in
Eq.~(\ref{eq:eom1}) in most cases below.

\begin{figure}[p]
\includegraphics[width=0.98\textwidth]{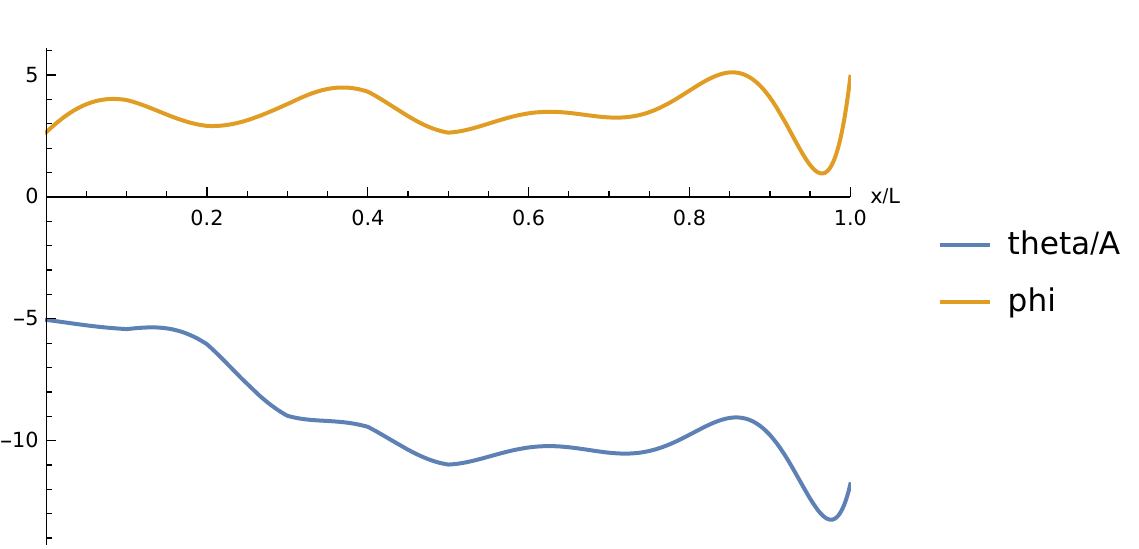}
\caption[...]{The angles $\theta(x)$ and $\phi(x)$ in the initial conditions in Eq.~(\ref{eq:M_x1}) that we obtain from a random
Wiener process. The angle $\theta(x)$ is multiplies with the amplitude $A=10^{-4}$.}
\label{fig:randini}
\end{figure}

To trigger flavor oscillations then requires small deviations from a pure flavor state as initial condition. For this we take
a random Wiener process between $x=0$ and $x=L$ for the angles in Eq.~(\ref{eq:M_x1}). The result is shown in Fig.~\ref{fig:randini}
where for the amplitude for $\theta(x)$ we take $A=10^{-4}$. This assures that initially the flavor state is very close
to the dominant flavor 1. A flavor-lepton number crossing with amplitude $b=0.5$ has been
chosen in Eq.~(\ref{eq:aniso2}) at $t=0$, whereas $a=0$, and thus no global anisotropy, has been assumed initially.
This flvaor number crossing, together with its shape at later times, will be shown further below in Fig.~\ref{fig:pol}.

In a supernova setting the various rates have the hierarchy $\lambda(x)\gtrsim\mu(x)\gg f_s(x)\simeq f_{\rm NC}(x)$.
In order to mimic this situation our parameter value choices will reflect this hierarchy.

We generally used $N_p=10$ and verified that for $N_p\gtrsim10$ the results do not depend significantly on the number of angular
modes $N_p$ any more. The simulations are performed with mathematica 12.1.

\begin{figure}[p]
%rand2_f3a
\includegraphics[width=0.48\textwidth]{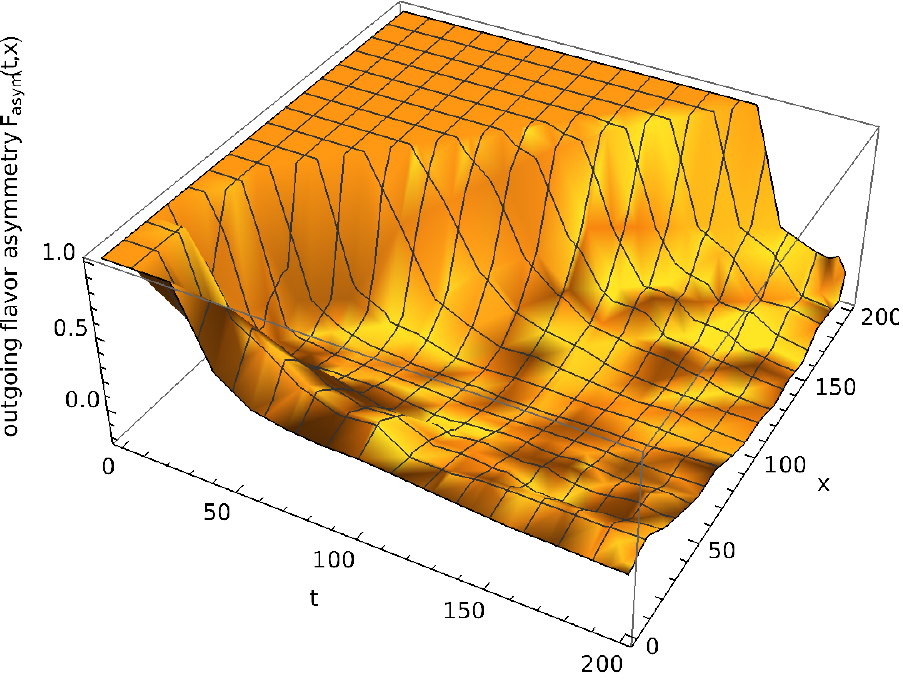}
\includegraphics[width=0.48\textwidth]{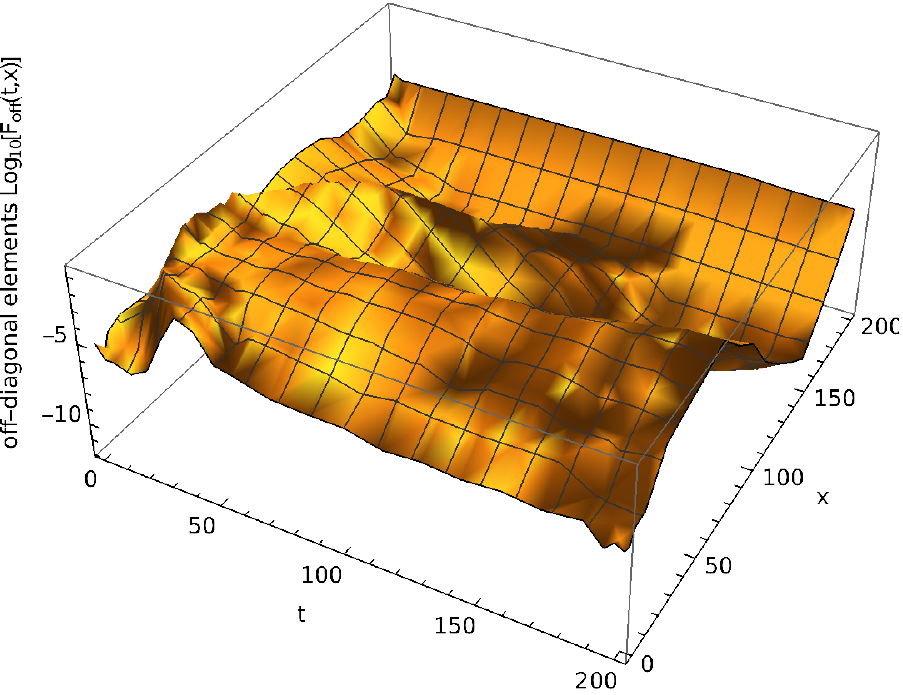}
\includegraphics[width=0.48\textwidth]{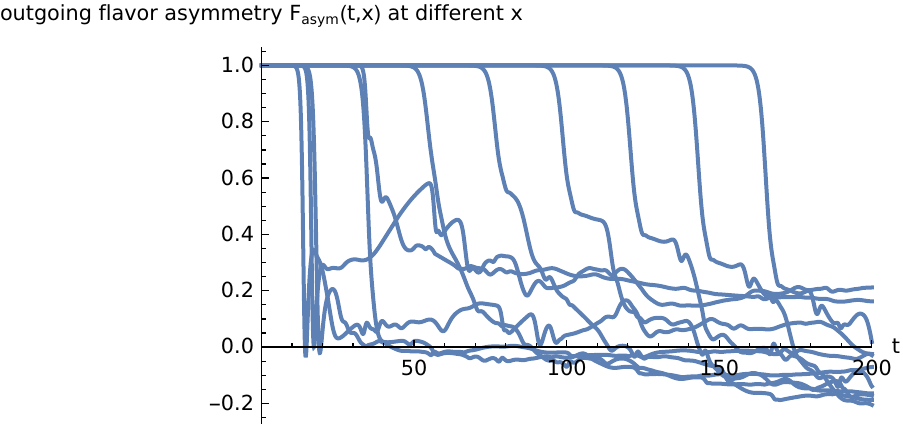}
\includegraphics[width=0.48\textwidth]{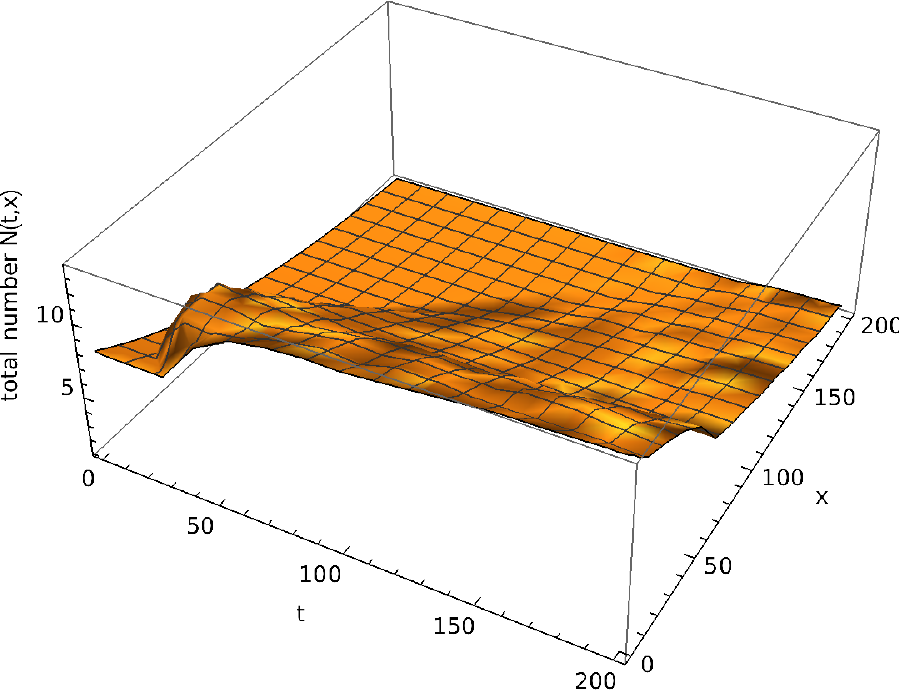}
\caption[...]{Results for a case (1) simulation with $N_p=10$ angular modes uniformly distributed in the range $0\leq v_x\leq1$, with
initially, at $t=0$, neutrinos and anti-neutrinos of mostly flavor 1
of total number $N(t=0,x)=N_pf_i(x)=0.8N_p\exp(-x/100)$. For $0<x<200=L$ random small deviations from a pure flavor 1 state have
been chosen, as described in the text. The total initial anisotropy is assumed to vanish, $a=0$, whereas a flavor crossing with $b=0.5$
was used Eq.~(\ref{eq:aniso2}). Further, there is no vaccum term and $\mu(x)=50\exp(-x/100)$, $\lambda(x)=0$,
with integration up to $t=200$. The charged current rates are proportional to $0.1\exp(-x/100)$, whereas there is no
neutral current scattering. Upper left: Normalized flavor asymmetry of outgoung modes defined in Eq.~(\ref{eq:F_asym}).
Upper right: Normalized off-diagonal elements defined in Eq.~(\ref{eq:F_off}). Lower left: Cuts through flavor asymmetry from upper left at
at 11 equidistant positions $x$ between $x=0$ and $x=200$. Lower right: Total number defined in Eq.~(\ref{eq:N_tot}).}
\label{fig:1}
\end{figure}

\begin{figure}[p]
%rand2_f3l2a
\includegraphics[width=0.48\textwidth]{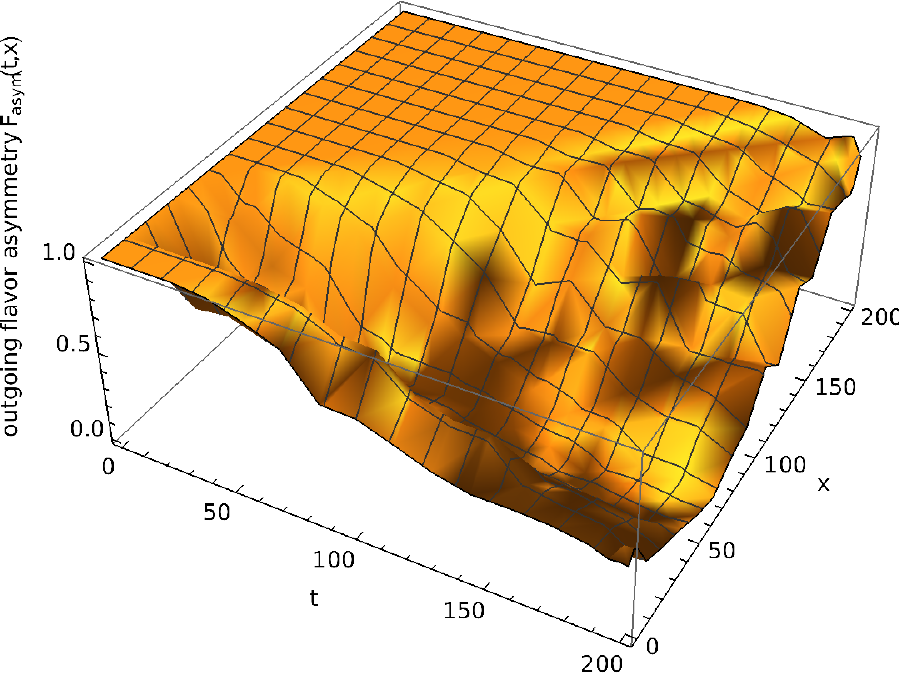}
\includegraphics[width=0.48\textwidth]{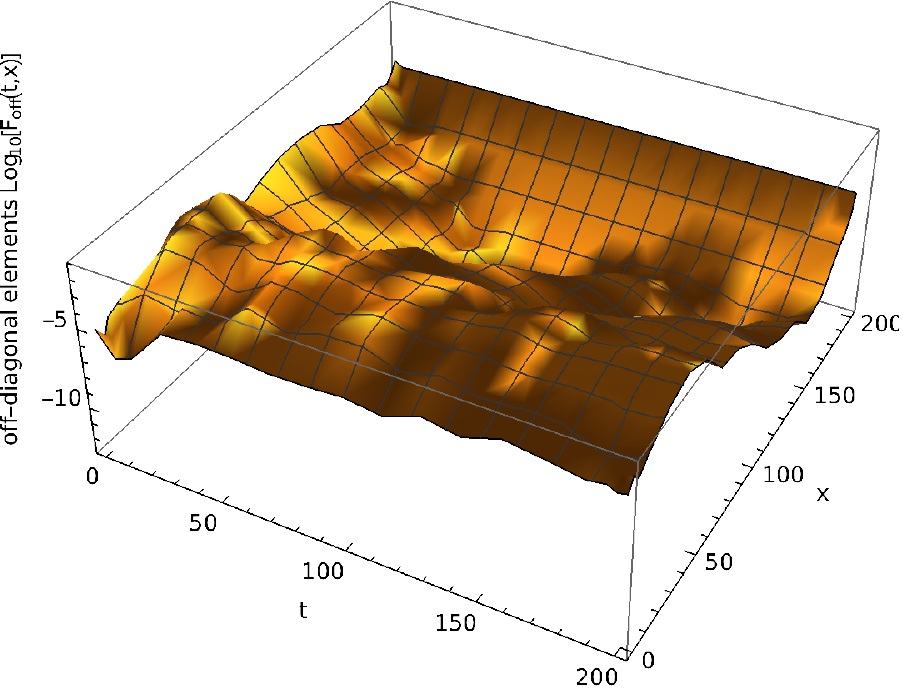}
\includegraphics[width=0.48\textwidth]{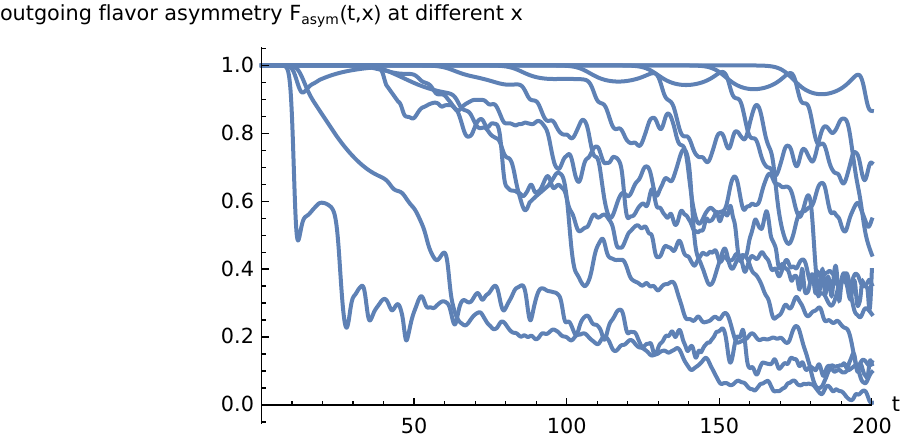}
\includegraphics[width=0.48\textwidth]{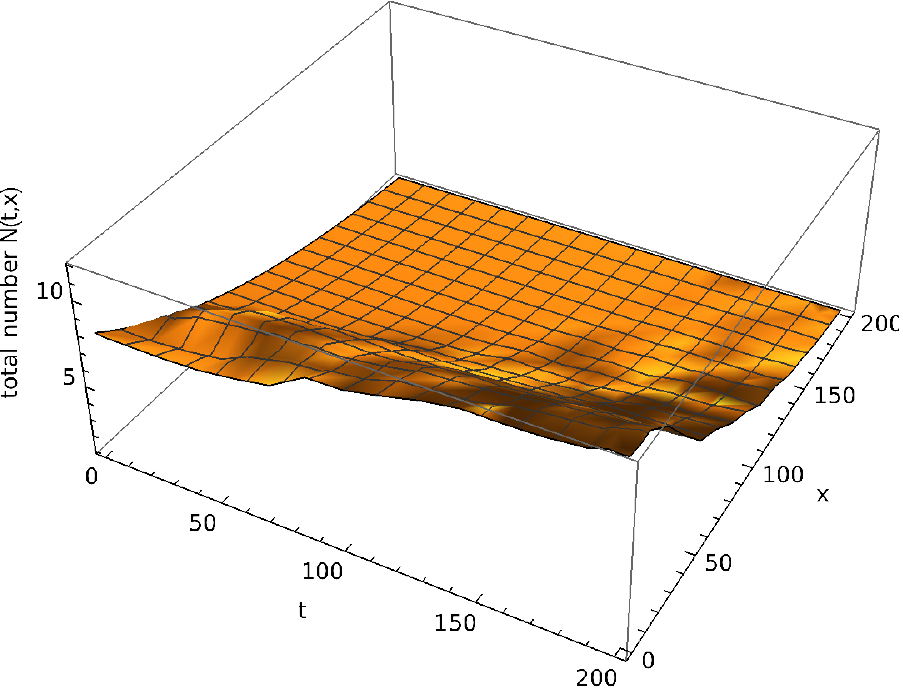}
\caption[...]{Results for a case (2) simulation which is identical to case (1) but with an addtional matter term
$\lambda(x)=\mu(x)=50\exp(-x/100)$. Note that the matter term partially suppresses and delays the flavor trsnsitions.}
\label{fig:2}
\end{figure}

We now simulated the following cases. In case (1) shown in Fig.~\ref{fig:1} we use $L=200$, $x_0=100$, a self-interaction
rate $\mu(x)=50\exp(-x/100)$,
no matter term, $\lambda(x)=0$, and no neutral current scattering terms, $f_{\rm NC}(x)=0$. Case (2) shown in Fig.~\ref{fig:2}
is identical to case (1) except that there is a matter term $\lambda(x)=\mu(x)=50\exp(-x/100)$. Case (3) has instead a
homogeneous matter term $\lambda(x)=50$. We also checked that choosing a constant $\mu(x)=50$ instead of the profile
does not change the results substantially.

\begin{figure}[p]
%rand2_f5
\includegraphics[width=0.48\textwidth]{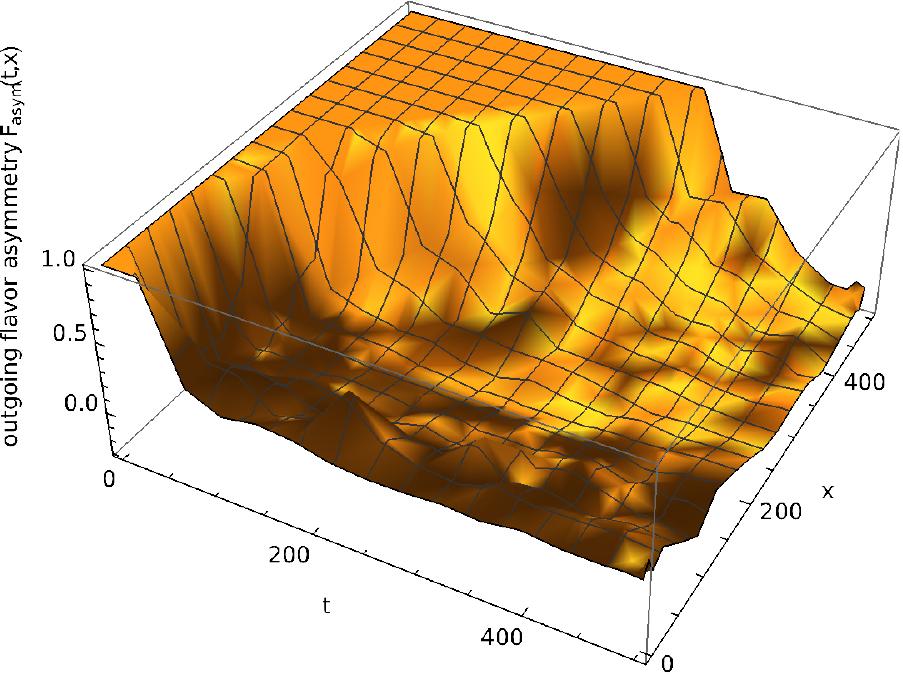}
\includegraphics[width=0.48\textwidth]{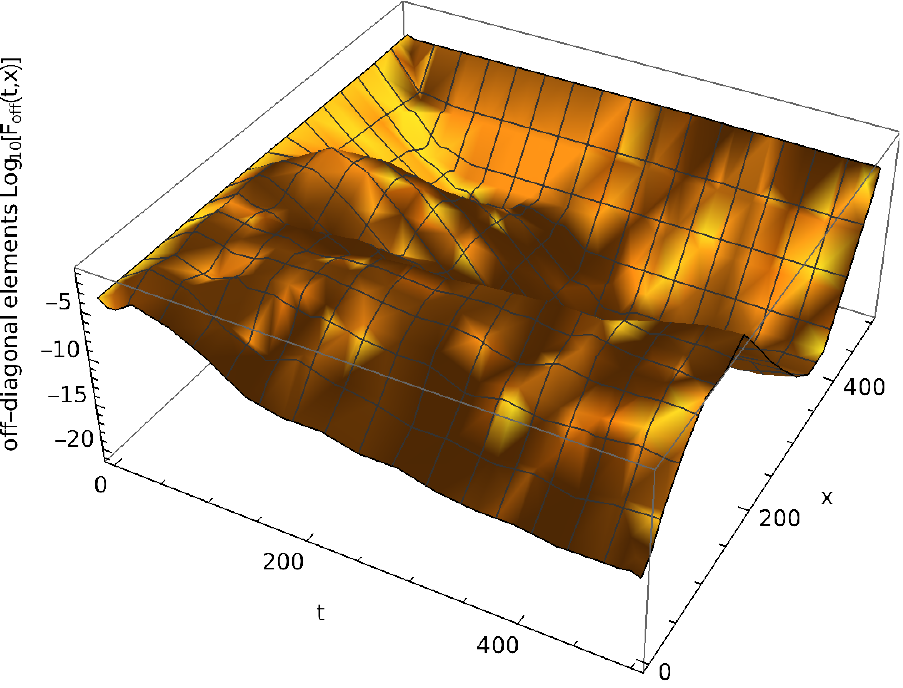}
\includegraphics[width=0.48\textwidth]{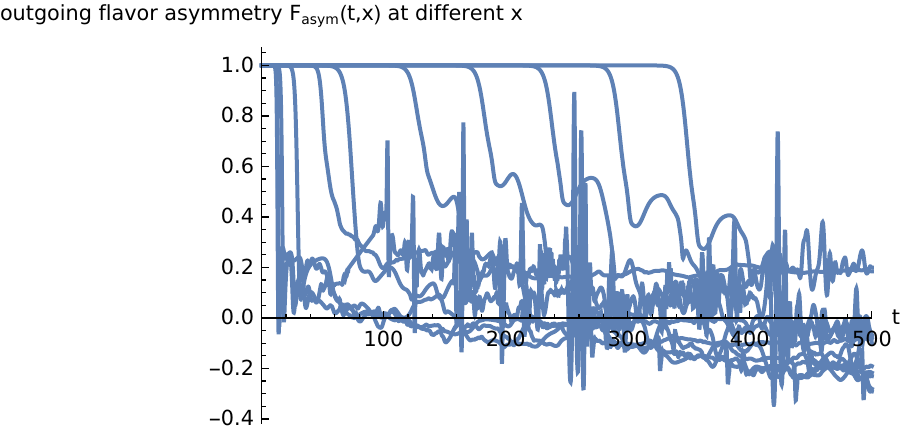}
\includegraphics[width=0.48\textwidth]{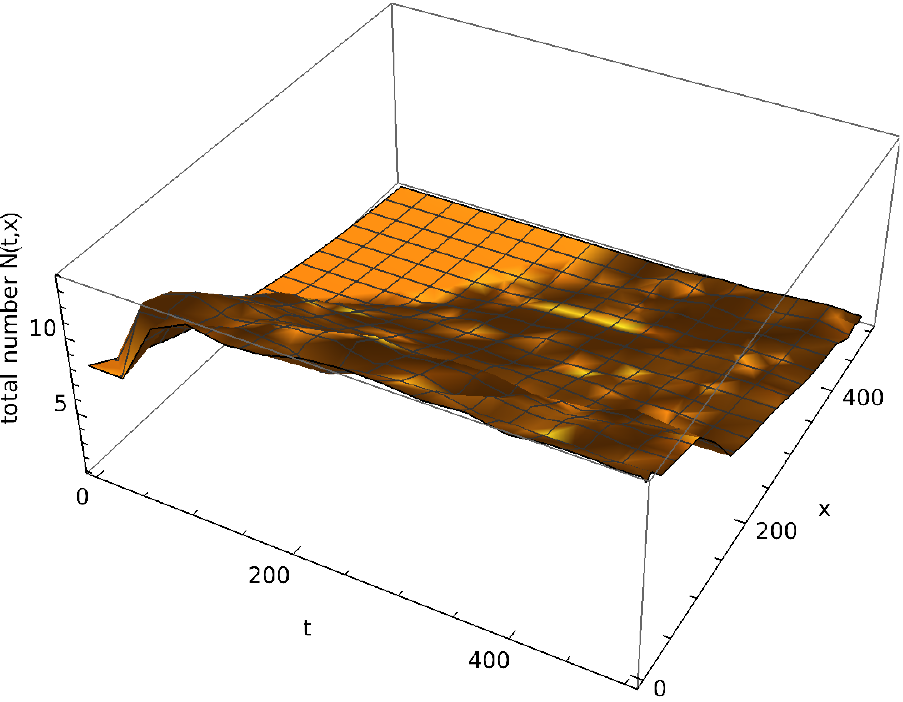}
\caption[...]{Results for a case (4) simulation which is very similar to case (1) shown in Fig.~\ref{fig:1}, except for shallower
profiles $\propto\exp(-x/250)$.}
\label{fig:4}
\end{figure}

\begin{figure}[p]
%rand2_f5la
\includegraphics[width=0.48\textwidth]{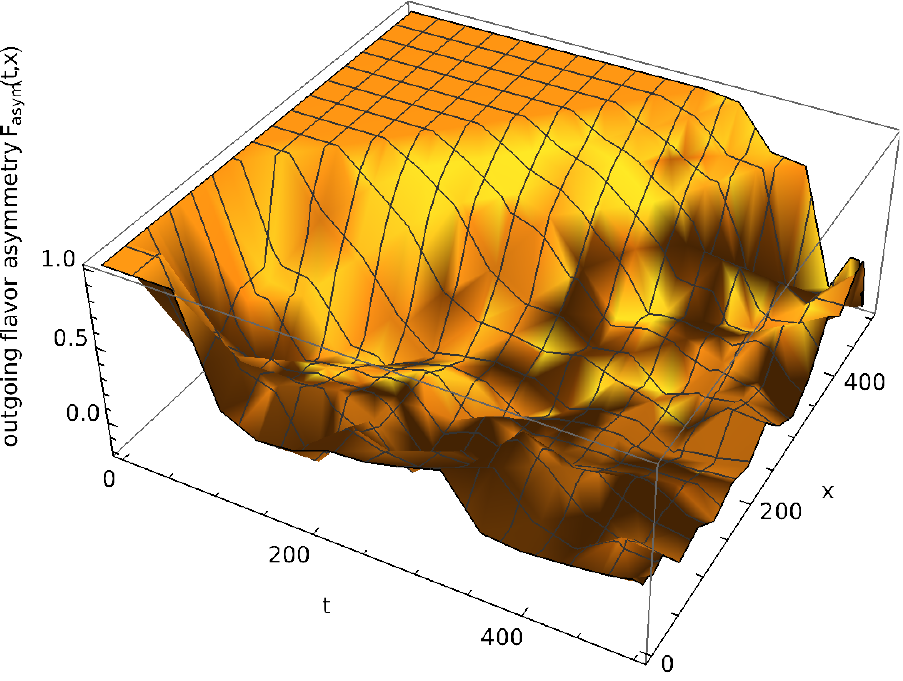}
\includegraphics[width=0.48\textwidth]{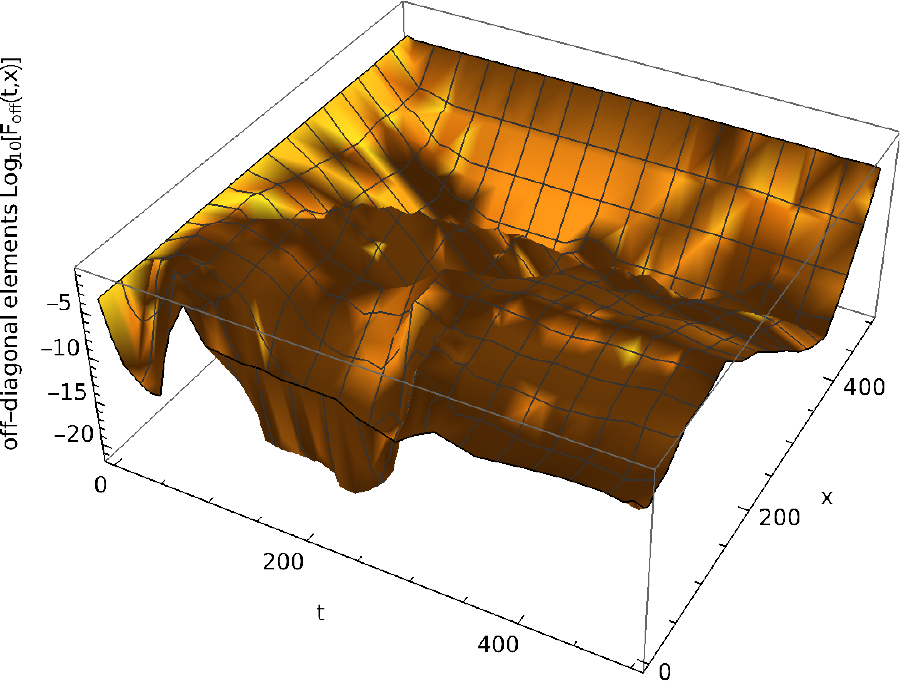}
\includegraphics[width=0.48\textwidth]{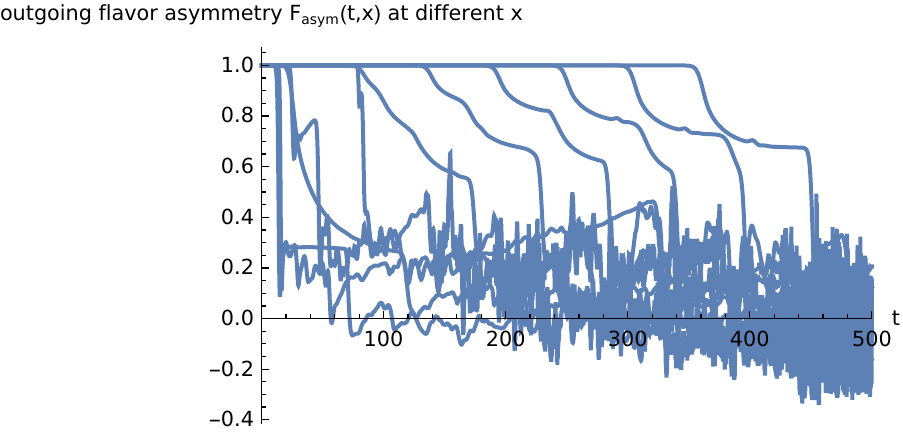}
\includegraphics[width=0.48\textwidth]{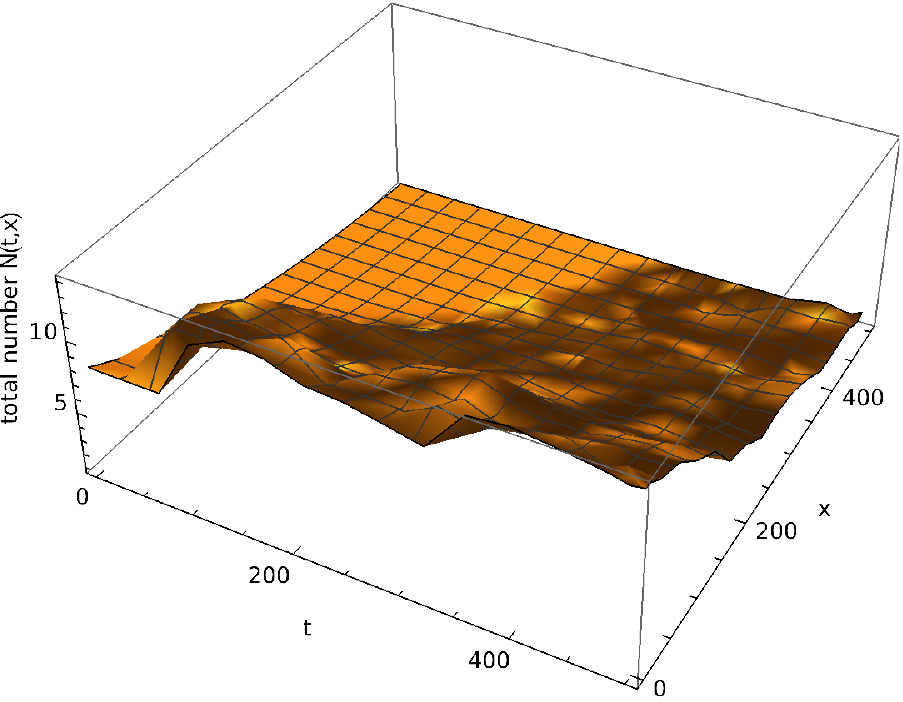}
\caption[...]{Results for a case (5) simulation which is identical to case (4) but with an addtional matter term
$\lambda(x)=\mu(x)=50\exp(-x/250)$. Note that the matter term partially suppresses and delays the flavor conversions.}
\label{fig:5}
\end{figure}

\begin{figure}[p]
%rand2_f5s
\includegraphics[width=0.48\textwidth]{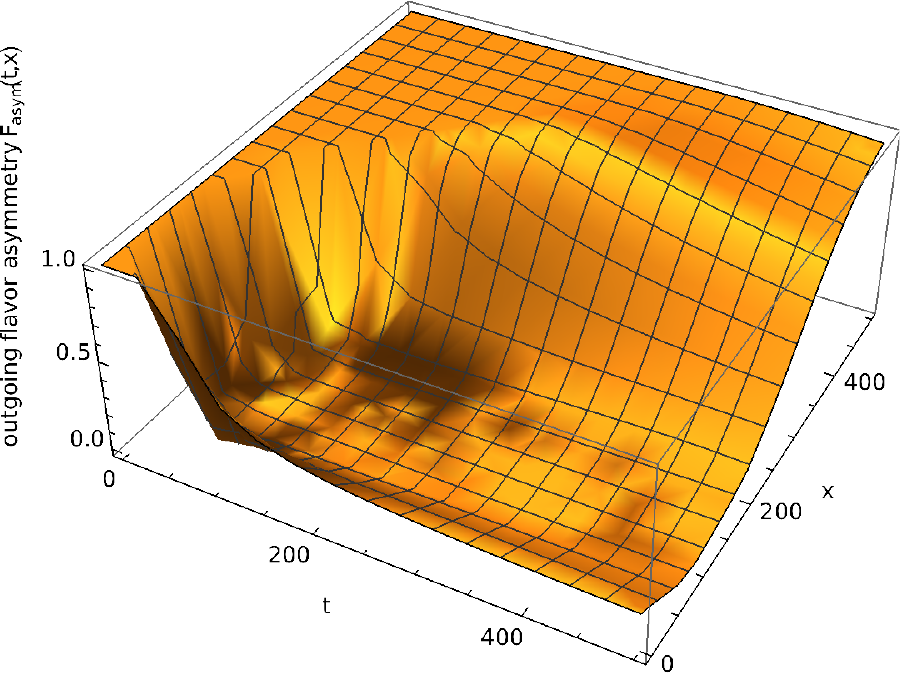}
\includegraphics[width=0.48\textwidth]{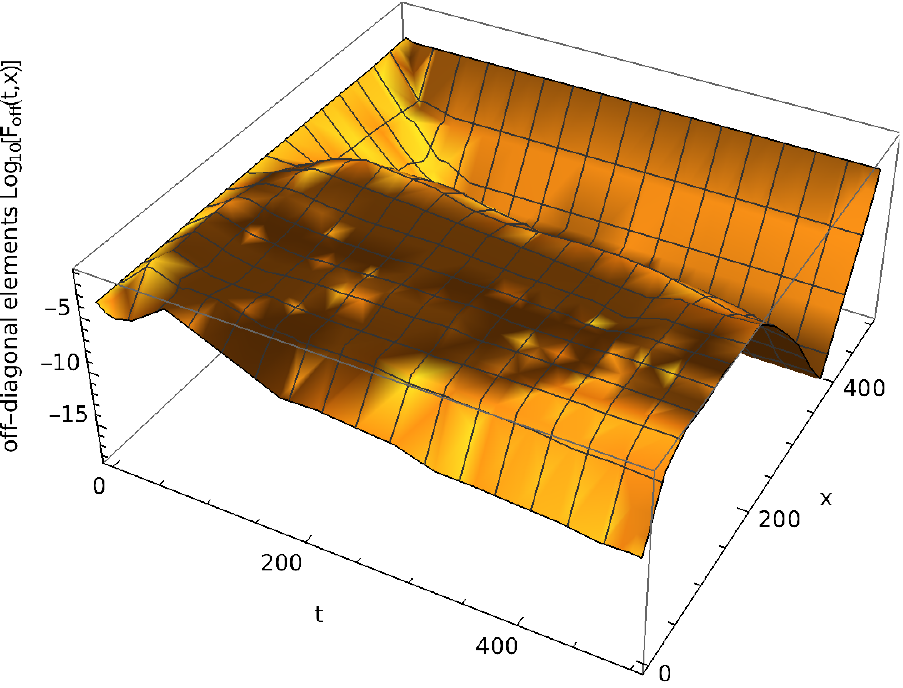}
\includegraphics[width=0.48\textwidth]{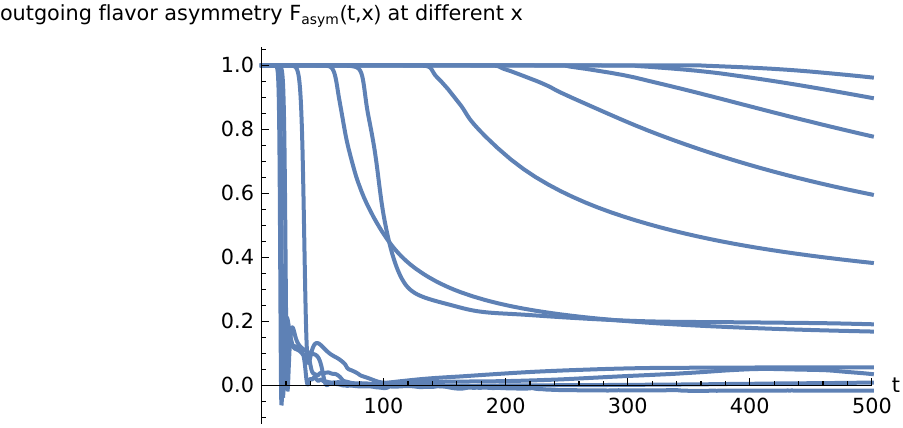}
\includegraphics[width=0.48\textwidth]{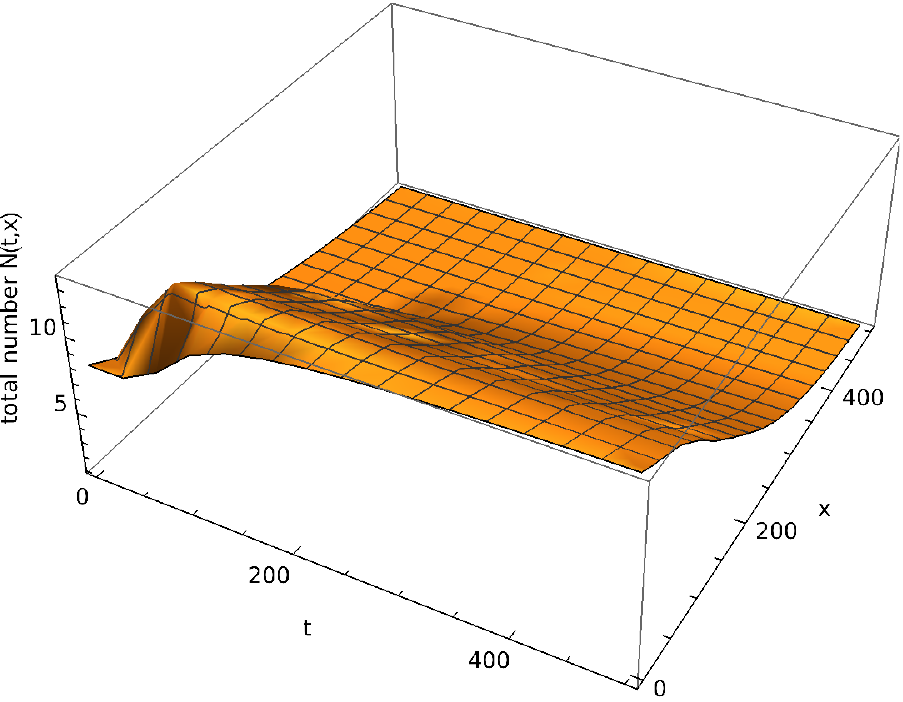}
\caption[...]{Results for a case (6) simulation which is identical to case (4) but with an addtional neutral current scattering term
$\propto f_{\rm NC}(x)=0.1\exp(-x/250)$. Note that the neutral current scattering term partially suppresses and delays the flavor conversions.}
\label{fig:6}
\end{figure}

\begin{figure}[p]
%rand2_f5sl
\includegraphics[width=0.48\textwidth]{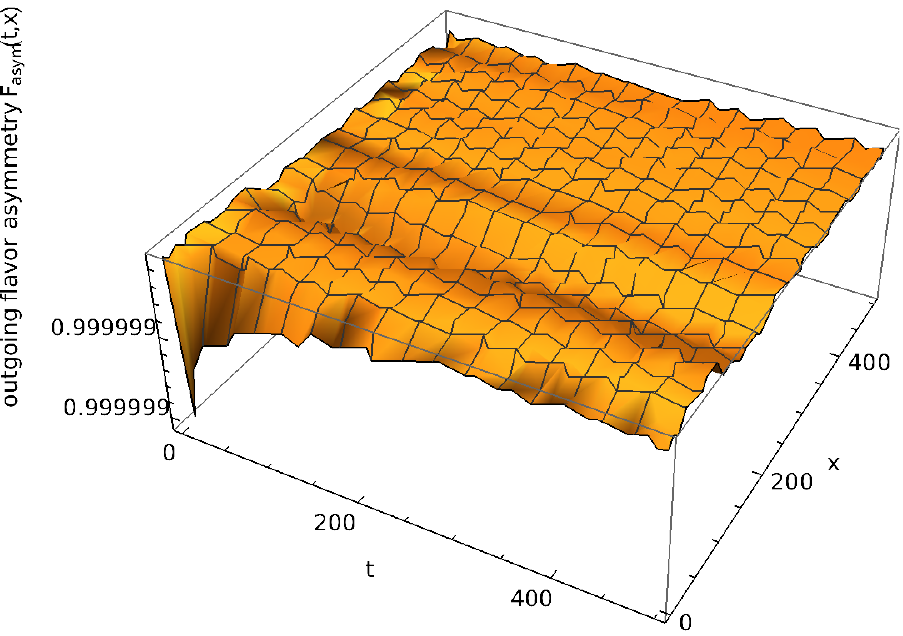}
\includegraphics[width=0.48\textwidth]{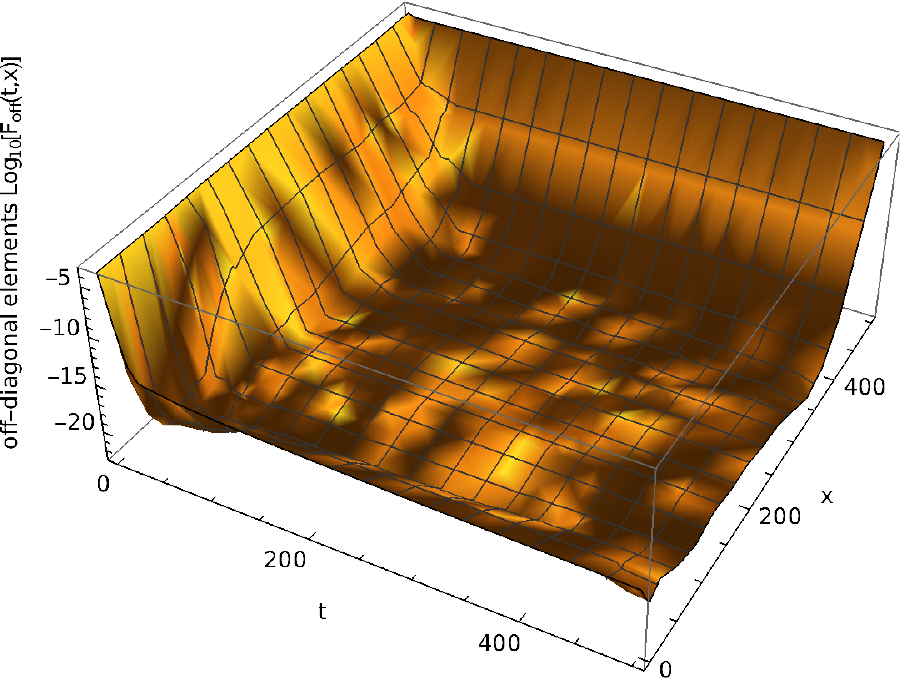}
\includegraphics[width=0.48\textwidth]{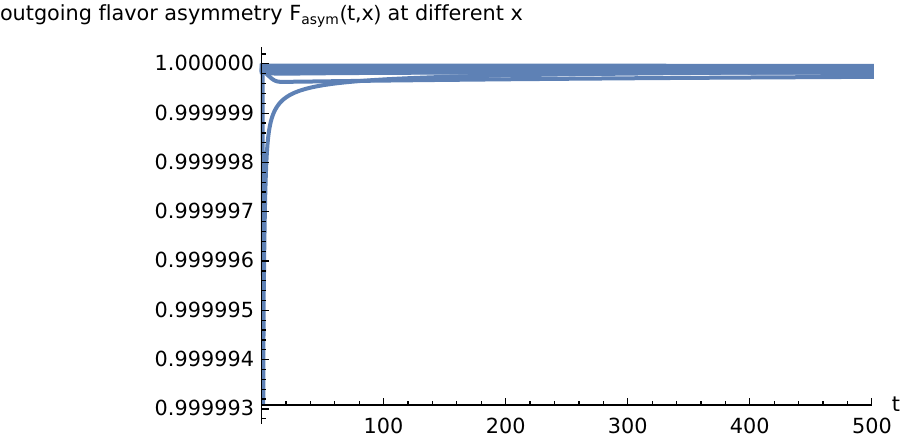}
\includegraphics[width=0.48\textwidth]{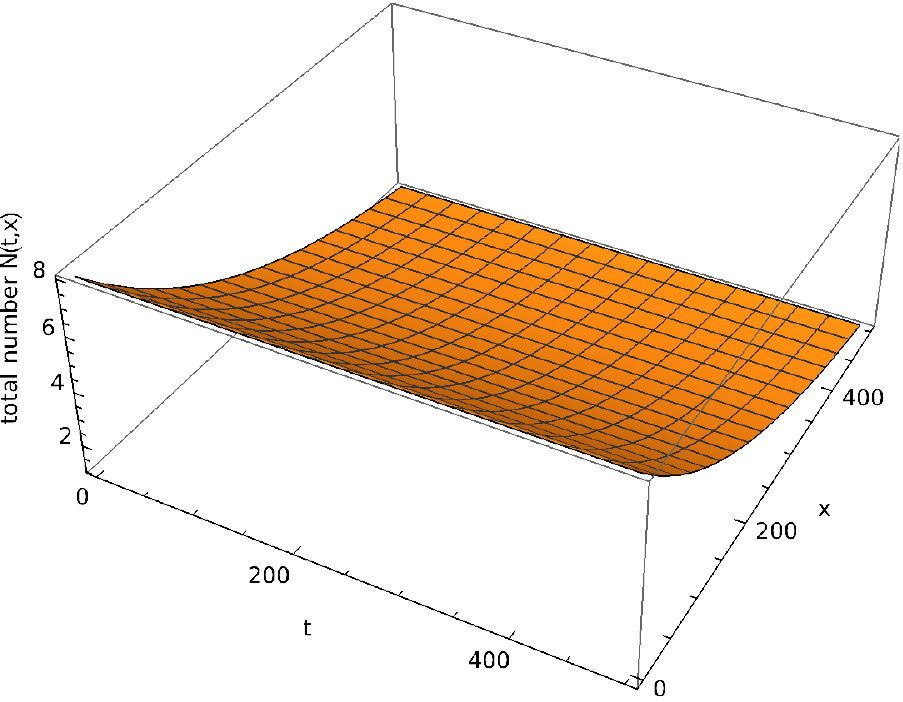}
\caption[...]{Results for a case (7) simulation which is identical to case (4) but with an addtional matter term
$\lambda(x)=\mu(x)=50\exp(-x/250)$ and an additional neutral current scattering term
$\propto f_{\rm NC}(x)=\exp(-x/250)$. Note that the flavor conversions are completely suppressed in this case.}
\label{fig:7}
\end{figure}

Next we investigate the influence of neutral current scattering and the interplay with matter terms. To this end
we use a shallower profile, $x_0=250$ with $L=500$. Case (4) shown in Fig.~\ref{fig:4} is identical to case (1)
apart from these changes. In case (5) shown in Fig.~\ref{fig:5} we add a matter term $\lambda(x)=\mu(x)=50\exp(-x/100)$,
whereas in case (6) shown in Fig.~\ref{fig:6} we have a neutral current scattering term with
$f_{\rm NC}(x)=0.1\exp(-x/250)$ in Eq.~(\ref{eq:kin_nc2}) instead. 
Finally, case (7) shown in Fig.~\ref{fig:7} combines a matter term $\lambda(x)=\mu(x)=50\exp(-x/100)$ with a
neutral current scattering term with normalisation $f_{\rm NC}(x)=\exp(-x/250)$.

\begin{figure}[p]
%rand2_f3l02a
\includegraphics[width=0.48\textwidth]{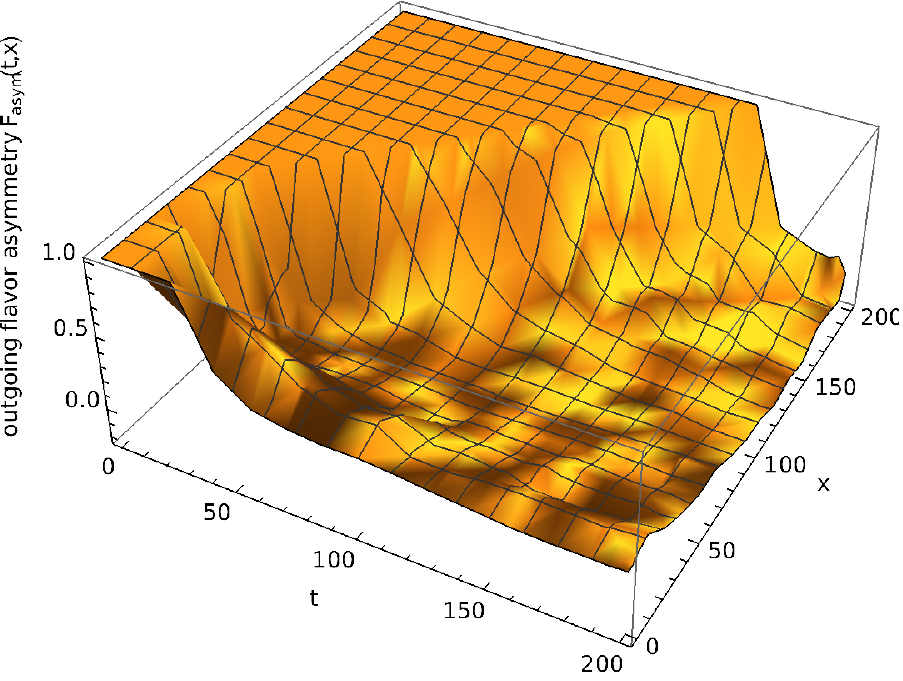}
\includegraphics[width=0.48\textwidth]{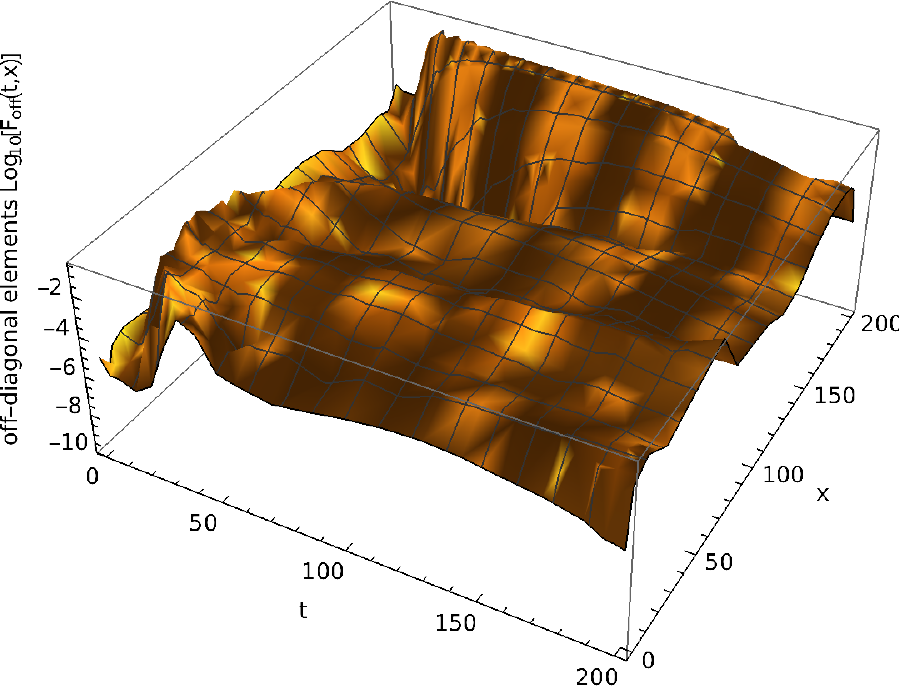}
\includegraphics[width=0.48\textwidth]{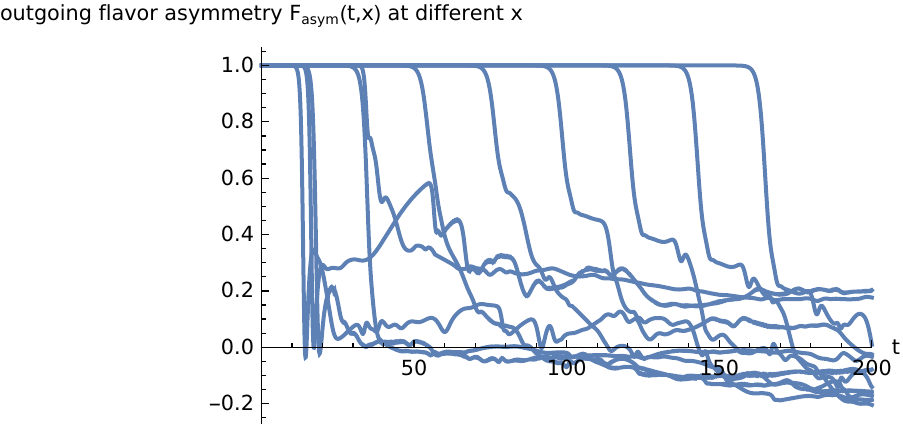}
\includegraphics[width=0.48\textwidth]{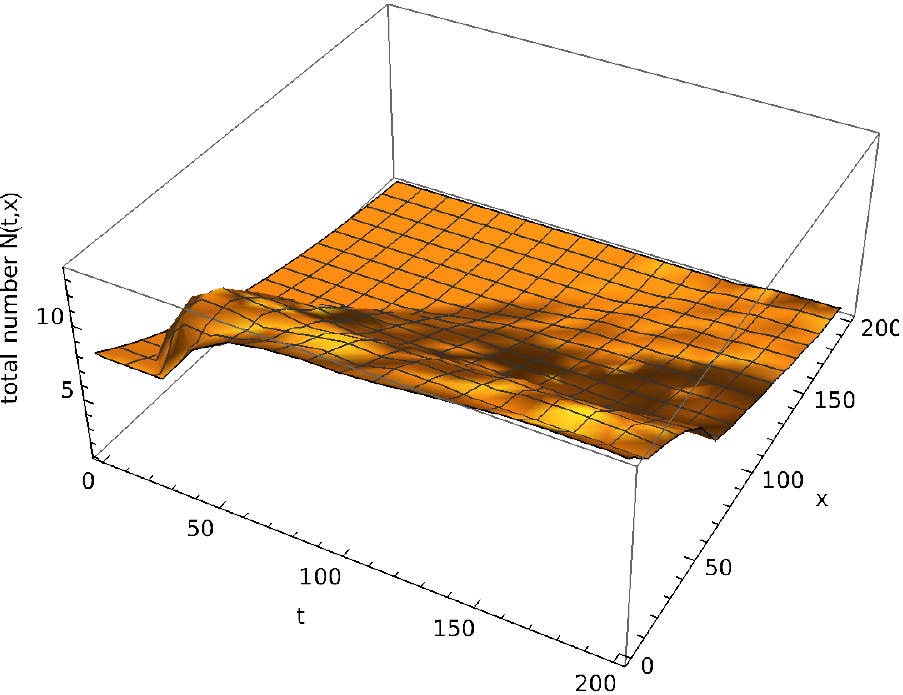}
\caption[...]{Results for a case (3) simulation which is identical to case (1) but with an addtional homogeneous matter term
$\lambda(x)=50$. Note that the matter term here has no discernible influence on the flavor evolution as the result for quantities
depending on flavor-diagonal terms is virtually identical to case (1) shown in Fig.~\ref{fig:1}, as expected on theoretical
grounds, see Sect.~\ref{sec:sec4}. However, the off-diagonal terms shown in the upper right paper are in fact different from case (1).}
\label{fig:3}
\end{figure}

In Fig.~\ref{fig:pol} we plot the normalized flavor-lepton number asymmetry corresponding to Eq.~(\ref{eq:ELN}),
\begin{equation}\label{eq:G}
  G(t,x,v_x)\equiv\frac{\rho_{11}(x,v_x)-\rho_{22}(x,v_x)-\bar\rho_{11}(x,v_x)+\bar\rho_{22}(x,v_x)}{N(t,x)}\,,
\end{equation}
as a function of direction $v_x=\cos\theta$ for cases (4) to (7), at several values for $t$ and $x$.
We see that both the matter term and the neutral current scattering term tend to suppress the initial asymmetry
already at early times which leads to a suppressed or delayed flavor conversions.

\begin{figure}[p]
\includegraphics[width=0.48\textwidth]{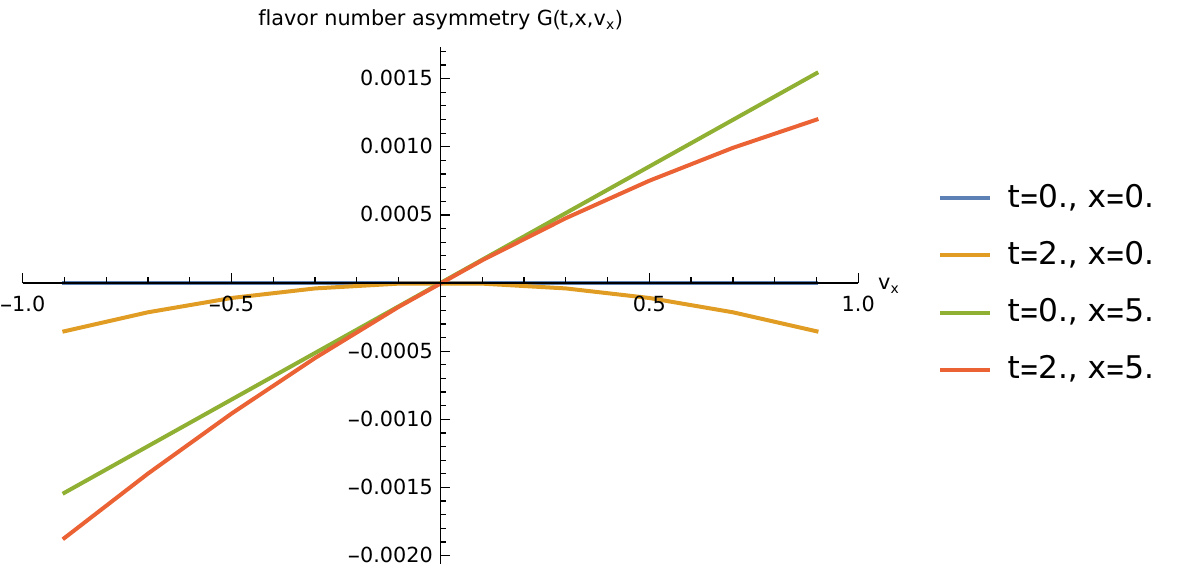}
\includegraphics[width=0.48\textwidth]{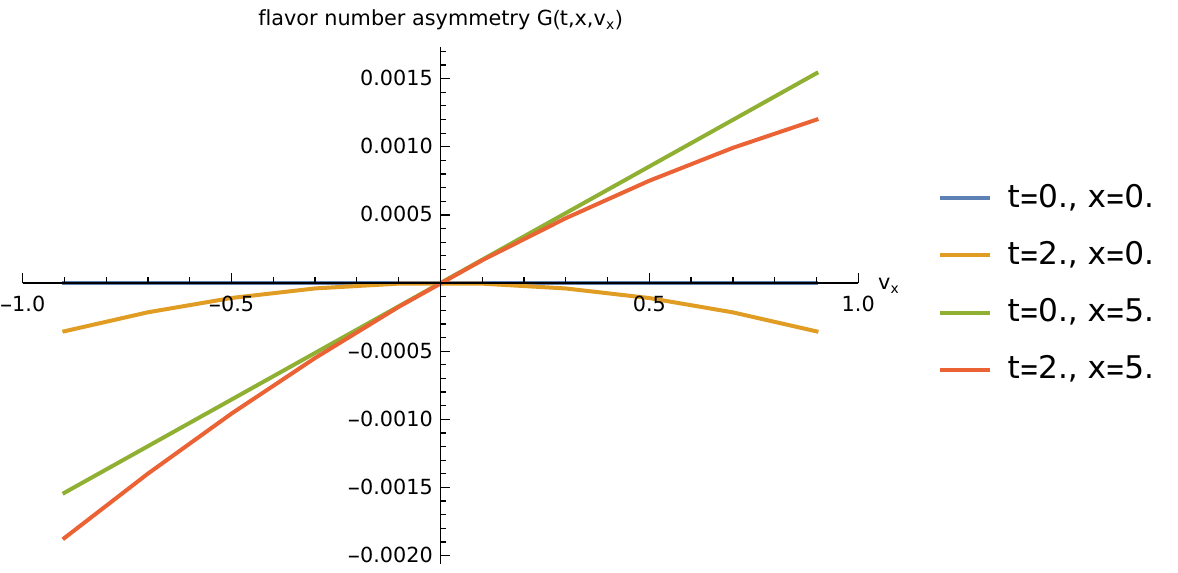}
\includegraphics[width=0.48\textwidth]{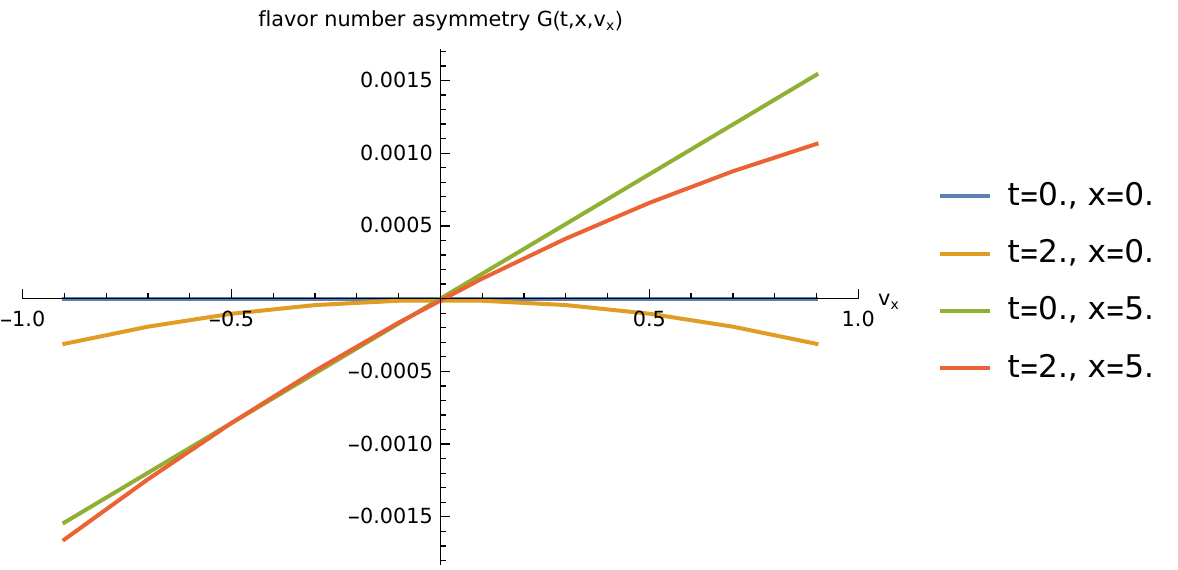}
\includegraphics[width=0.48\textwidth]{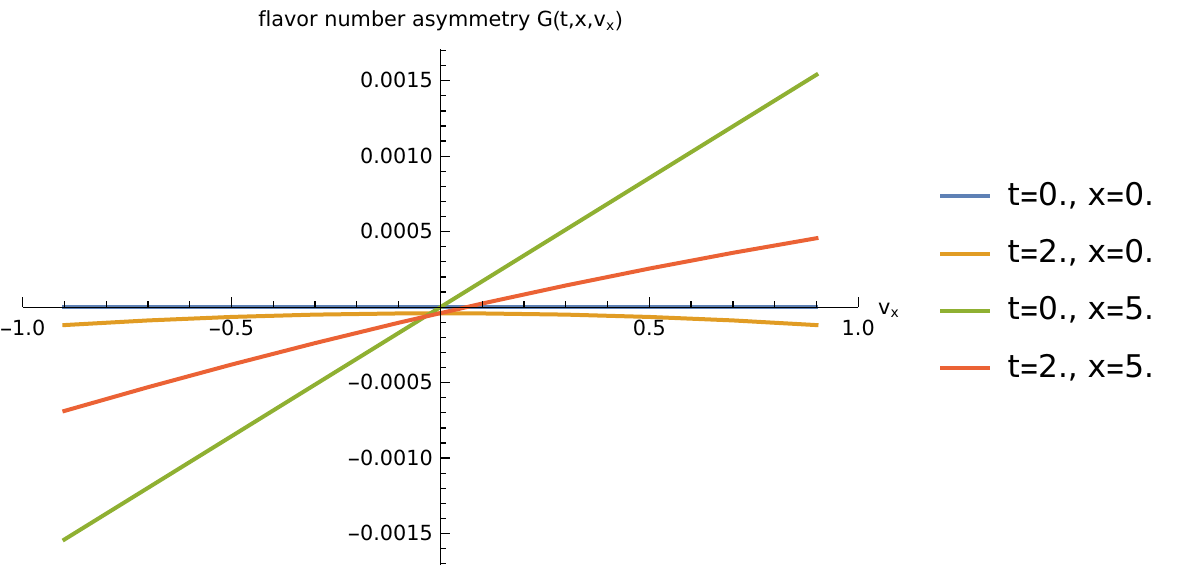}
\caption[...]{The normalized flavor-lepton number asymmetry defined in Eq.~(\ref{eq:G}) as a function of direction $v_x$
for cases (4) (upper left), (5) (upper right), (6) (lower left) and (7) (lower right), at the values for $t$ and $x$ indicated.
Note that at $t=0$ the flavor-lepton number asymmetry is of order $bg(x)/N_p$ which for $x=5$ gives $G(t=0,x=5,v_x)\simeq0.00154v_x$.
At $x=L/2$ (not shown) the initial flavor asymmetry becomes maximal, $G(t=0,x=L/2,v_x)\simeq0.018v_x$.
Note that both the matter term and the neutral current scattering term tend to suppress the asymmetry already at early times.}
\label{fig:pol}
\end{figure}

\begin{figure}[p]
%rand2_f52a
\includegraphics[width=0.48\textwidth]{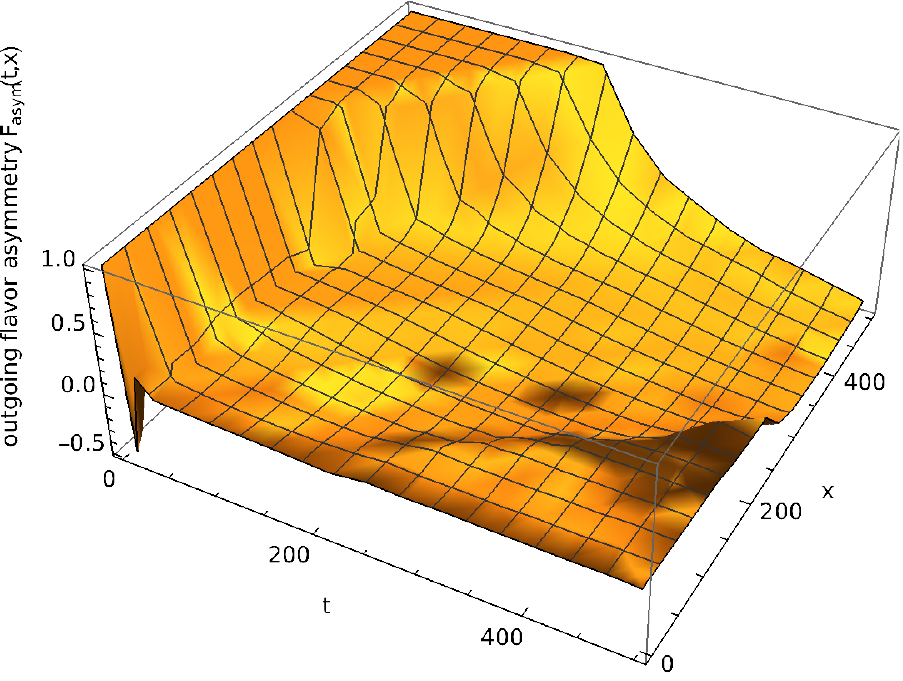}
\includegraphics[width=0.48\textwidth]{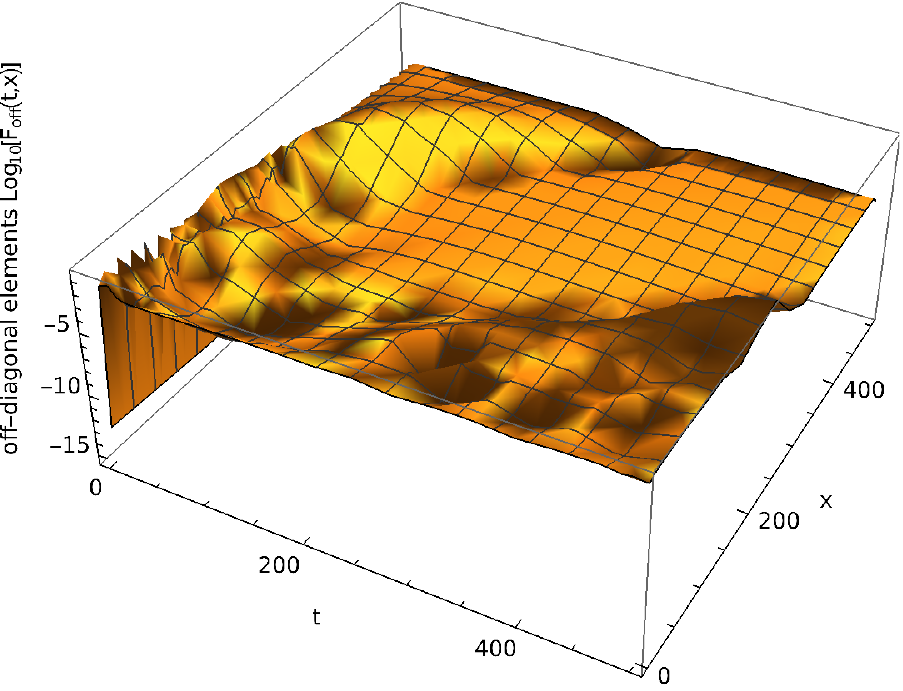}
\includegraphics[width=0.48\textwidth]{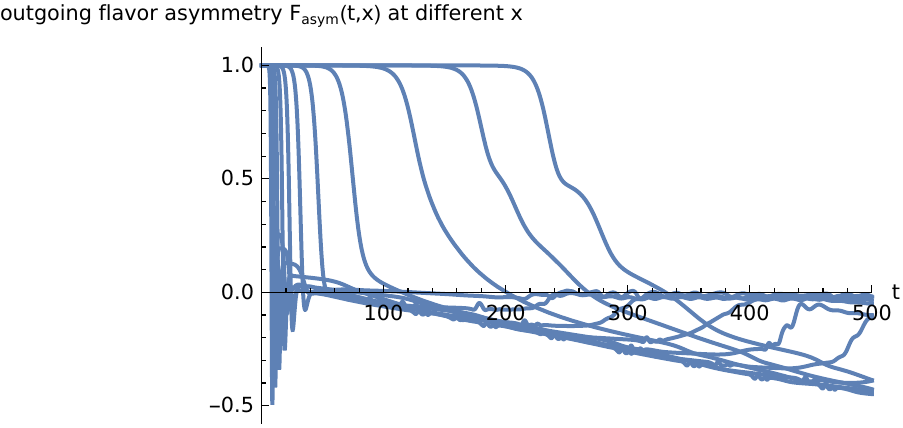}
\includegraphics[width=0.48\textwidth]{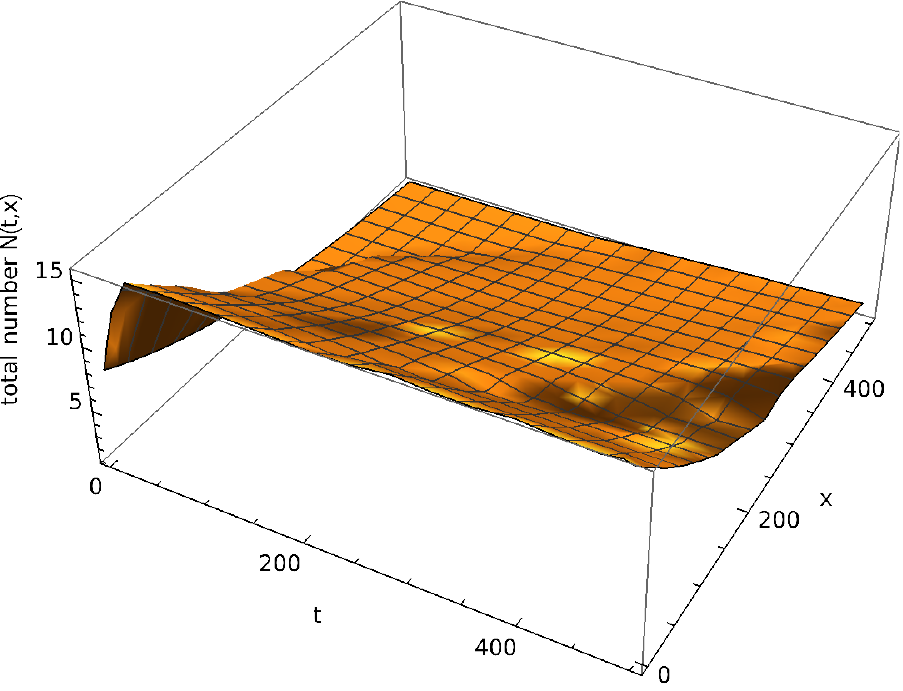}
\caption[...]{Results for case (8) simulation which represents a slow flavor conversion scenario starting with an inital pure
flavor state and inverted hierarchy vacuum terms with $\Delta m^2=0.1$, $\theta_0=10^{-4}$, an otherwise identical parameters to case (4).}
\label{fig:8}
\end{figure}

Finally, in case (8) we consider for comparison a scenario corresponding to so-called slow flavor conversions.
In this case, we start with initial conditions for a pure flavor 1, $A=0$, and thus $M(x)={\rm diag}(1,1)$ in
Eqs.~(\ref{eq:M_x3}) and~(\ref{eq:M_x1}). The flavor conversions are instead triggered by a non-vanishing vacuum
term with $\Delta m^2=0.1$, $\theta_0=10^{-4}$, i.e. an inverted hierarchy with a small vacuum mixing angle. All
other parameters are as in case (4). The results are shown in Fig.~\ref{fig:8}. In contrast to the previous cases one can
see bipolar oscillations at small times in which the flavor asymmetry goes considerably below zero.

%From these examples one can derive some general tendencies: A large $\lambda(x)$ seems to strongly suppress transitions
%for a steep $x-$profile, but not for a flat/constant profile.
In the following section we discuss these results in more detail.

\section{Discussion and Outlook}\label{sec:sec4}
Let us first mention a few general aspects. It can be seen in the lower right panels of Figs.~\ref{fig:1} to~\ref{fig:7}
that conversions from the originally dominating flavor 1 locally leads to an increase of the number of neutrinos,
or neutrino density, at moderate radii. This is easy to understand since a decrease of flavor 1 occupation numbers leads to
production of flavor 1 states by the charged current interactions. On the other hand, at large radii, close to the outer
boundary, the neutrino density changes at most due to additional neutrinos produced in the inner region streaming
outward. This is because at large radii the charged current interactions are too small to significantly change the
neutrino number. Thus, the range of radii simulated in our examples indeed effectively covers regions within and
outside the neutrino sphere.

The momentum sum of the off-diagonal terms shown in the upper right panels of Figs.~\ref{fig:1} to~\ref{fig:7}
typically show a rise in the regions where flavor conversions take place, but tend to get strongly suppressed in regions
where no collective oscillations happen, as one would expect. An exception is the outer boundary at $x=L$ where the off-diagonal
components summed over all momenta (in- as well as outgoing) are essentially fixed by the boundary
condition for modes propagating inwards Eq.~(\ref{eq:M_init2a}) which equals the initial condition
of flavor states close to flavor 1. This implies that the outgoing modes have very small off-diagonal elements,
probably because they are averaged out by the oscillations and convective transport. Note that 
the upper left panels of Figs.~\ref{fig:1} to~\ref{fig:7} show the flavor asymmetry of only the outgoing modes,
see Eq.~(\ref{eq:F_asym}), whereas the off-diagonal elements shown in the upper right panels are summed over all modes,
see Eq.~(\ref{eq:F_off}). Also note that efficient fast flavor conversions always saturate near flavor equilibration.

Using stability analysis for linear perturbations in a homogenous system characterized by $\mu\equiv\mu(x)=$const
one can show that fast flavor conversion instabilities occur for
\begin{equation}\label{eq:stab1}
  |k_x|,\frac{1}{\sigma}<k_c\simeq\frac{4\mu b}{|v_1-v_2|}\,,
\end{equation}
and with a growth rate
\begin{equation}\label{eq:stab2}
  R\simeq2\mu b\,,
\end{equation}
see e.g. Ref.~\cite{Chakraborty:2016yeg}.
%formuli to be checked
The results shown in Figs.~\ref{fig:1} to~\ref{fig:5} appear roughly consistent with this in the linear growth phase when
substituting $\mu(x)bg(x)\simeq\mu_0\sin(\pi x/L)\exp(-2x/x_0)$ for $\mu b$.

Furthermore, the upper left panels of Figs.~\ref{fig:1} to~\ref{fig:7} show that
both neutral current scattering terms and refractive effects from the matter tend to partially suppresses or at least delay
the flavor conversions. Let us now discuss the role of these two ingredients in more detail.

In general, if the matter term in Eq.~(\ref{eq:Omega_m}) is independent of the spatial coordinates, $\Omega_m({\bf r})=\lambda D$
with $D$ a constant diagonal matrix with order unity entries and $\lambda$ a constant rate, one can define modified density matrices through
\begin{equation}
  \tilde\rho(t,{\bf r},{\bf p})\equiv\exp\left[+i\lambda Dt\right]\rho(t,{\bf r},{\bf p})\exp\left[-i\lambda Dt\right]\,,
\end{equation}
and an analogous equation for $\tilde{\bar\rho}(t,{\bf r},{\bf p})$ for the anti-neutrinos.
Since $D$ commutes with $\Omega_m$, $G_{\rm S}$, $G$ and the charged current matrix
rates ${\cal P}({\bf r},{\bf p})$ and ${\cal A}({\bf r},{\bf p})$ in Eq.~(\ref{eq:kin_cc}, it is easy to see that if
$\rho(t,{\bf r},{\bf p})$ and $\bar\rho(t,{\bf r},{\bf p})$ obey Eq.~(\ref{eq:eom1}), then $\rho(t,{\bf r},{\bf p})$ and
$\tilde{\bar\rho}(t,{\bf r},{\bf p})$ also obey Eq.~(\ref{eq:eom1}) but without the matter term $\Omega_m$, as long
as there is no off-diagonal vacuum term $\Omega^0_{\bf p}$ with which $D$ does not commute. In this
sense it is often said that the matter term is ``rotated away'' and thus effectively eliminated from the problem as long
as one is mostly interested in the diagonal (flavor) content of the density matrices, see e.g. Ref.~\cite{Chakraborty:2016yeg}.
This is confirmed by a comparison of cases (1) and (3) shown in Figs.~\ref{fig:1} and~\ref{fig:3}, respectively.
The influence of a homogeneous matter term is very small even in the presence of vacuum terms
as long as the matter term is much larger than the
vacuum term, $\lambda\gg\Delta m^2/(2p)$, whose off-diagonal components then tend to average out~\cite{Duan:2005cp}.

We note that some works in the literature did find an influence of even a constant matter
term on collective oscillations. For example, in Ref.~\cite{Esteban-Pretel:2008ovd} it was found
that electron densities larger than the neutrino densities can lead to multi angle decoherence
which tends to suppress collective flavor conversions. Given the above analytical argument this
may seem surprising, however those findings have usually been made assuming stationary
solutions which turns the problem into an ordinary differential equation in the radial coordinate.
To directly put this in relation with the partial differential equations considered in the present
work would require to adopt time independent boundary conditions. In case of slow collective
oscillations in the presence of vacuum mixing those would typically be given by a pure
flavor state. In case of fast collective oscillations without vacuum terms the boundary conditions
would be almost pure flavor states with small off-diagonal terms that are time independent
in the standard frame, and rotating with frequency $\lambda$ in the rotating frame. In the latter
case this induces a $\lambda$ dependence after all, even though $\lambda$ can still be eliminated
from the partial differential equation. This is
quite different from the boundary conditions we use here and which in particular allow non-trivial
time evolution at the boundaries. The case of stationary solutions can, therefore, not directly
be compared with the time dependent scenarios of our present work. This also shows us that the
solutions can significantly depend on the chosen boundary conditions.

Coming back to the Liouville-type equation with collision terms Eq.~(\ref{eq:Omega_m}), it is less clear if the matter
term can also be rotated away in case of a spatially inhomogeneous matter term.
Naively one might expect that this should still be a reasonably good approximation as long as the spatial scale
$d$ on which the rates entering the problem vary satisfies $\lambda({\bf r})\gg1/d$ in the sense that the variations 
are then adiabatic. Our simulations show, however, that even in the case $\lambda({\bf r})\sim50$, $d\sim250$, i.e.,
$\lambda({\bf r})\sim10^4/d$ and in the absence of vacuum terms the matter term
has a discernible effect and tends to slow down fast flavor conversions. This can be seen by comparing
Fig.~\ref{fig:1} with Fig.~\ref{fig:2} and Fig.~\ref{fig:4} with Fig.~\ref{fig:5} . It is thus possible that small scale 
variations of rates, induced for example by turbulent motion, may have a significant effect on collective flavor 
oscillations.

We also found significant effects of neutral current scattering on flavor evolution.
In contrast to Ref.~\cite{Shalgar:2020wcx,Sasaki:2021zld} in our examples non-forward scattering of neutrinos does not enhance fast 
flavor conversions. This may have to do with the fact that Ref.~\cite{Shalgar:2020wcx,Sasaki:2021zld} considered a homogeneous 
system, reducing the system to an ordinary differential equation. In Fig.~\ref{fig:6} we see that when neutral and 
charged current interactions occur with comparable rates, significant flavor conversions only take place at small radii 
where $\mu(x)$ is large. In addition, those conversions do not propagate to the outer boundary, so that the neutrino 
flux leaving the system is essentially still of pure flavor 1. Isotropic scattering tends to smooth out the crossing of the 
flavor-lepton number asymmetry which drives fast oscillations, as can be seen in Fig.~\ref{fig:pol}. This is the main 
reason why scattering tends to suppress fast flavor conversions.
Interestingly, we found that if one increases the neutral current scattering rate by a factor 10 (not shown here)
the isotropization becomes so efficient close to $x=0$ that neutrinos stay essentially in the flavor 1 state even around
the inner boundary.

Finally, when both a matter term and non-forward neutral current scattering are present simultaneously, any fast
flavor conversions seem to be completely prohibited, as is seen in Fig.~\ref{fig:7}.

Our framework also allows to simulate other cases such as density matrices for more than two flavors, inclusion of momentum
modes with different energies and a realistic implementation of neutrino pair processes due to neutral current
interactions, as well as cases with a global asymmetry between neutrinos and anti-neutrinos.
We leave a more systematic investigation of such cases and a more detailed discussion of so-called slow flavor
conversions in the presence of vacuum oscillation terms, in particular in the inhomogeneous settings studied here, to future work.

\section{Conclusions}\label{sec:sec5}
In this paper we have simulated collective neutrino oscillations in the context of inhomogeneous rates for the
forward-scattering self-interactions and matter potential, as well as for charged current production
and absorption and neutral-current non-forward scattering. Those rates were chosen with a hierarchy that mimics
the case of a core collapse supernova close to the neutrino sphere and profiles that fall off with increasing radius.
We found a considerable influence of inhomogeneous matter induced refractive terms and neutral current non-
forward scattering on collective neutrino oscillations with a tendency to suppress or delay in particular fast flavor 
conversions, in particular if both are present. Furthermore, solutions can depend significantly on the boundary
conditions. We do not pretend that the simulations performed here are directly 
applicable to the situation of a core collapse supernova.
However, in our opinion these findings should serve as a warning that such effects should be taken into account in 
any realistic treatment of collective neutrino oscillations.

\begin{acknowledgments}
G.S. acknowledges support by the Deutsche Forschungsgemeinschaft (DFG, German Research Foundation) under Germany’s Excellence Strategy — EXC 2121 “Quantum Universe” — 390833306. We also acknowledge numerous insightful discussions on the subject of
collective neutrino oscillations with Sajad Abbar, Francesco Capozzi, Basudeb Dasgupta, George Fuller, Alessandro Mirizzi, Georg Raffelt,
Shashank Shalgar, Irene Tamborra, and many others.
\end{acknowledgments}

%\bibliography{../bib}

\bibliography{flavor_simulations}

\end{document}